\documentclass[letterpaper,longbibliography,american,prx,twocolumn,amsmath,amssymb,superscriptaddress,nofootinbib]{revtex4-2}
\usepackage[T1]{fontenc}
\usepackage{babel}
\usepackage{prettyref}
\usepackage{float}
\usepackage{mathtools}
\usepackage{dsfont}
\usepackage{amsmath}
\usepackage{amssymb}
\usepackage{graphicx}
\usepackage{cancel}
\usepackage[unicode=true,
 bookmarks=true,bookmarksnumbered=false,bookmarksopen=false,
 breaklinks=false,pdfborder={0 0 1},backref=false,colorlinks=false]
 {hyperref}
\hypersetup{pdftitle={Learning actions from Data using Invertible Neural Networks},
 pdfauthor={Claudia Merger, Alexandre Rene, Kirsten Fischer, Peter Bouss, Sandra Nestler, David Dahmen, Carsten Honerkamp, Moritz Helias},
 colorlinks=true,linkcolor=black,citecolor=black,urlcolor=black,filecolor=black}

\makeatletter

\pdfpageheight\paperheight
\pdfpagewidth\paperwidth

\usepackage{tikz}
\usetikzlibrary{calc}

\usepackage[latin1]{inputenc}
\usepackage{amsmath}
\newrefformat{eq}{\hyperref[#1]{Eq.~(\ref{#1})}}
\newrefformat{cap}{\hyperref[#1]{Fig.~\ref{#1}}}
\newrefformat{fig}{\hyperref[#1]{Fig.~\ref{#1}}}
\newrefformat{tab}{\hyperref[#1]{Table ~\ref{#1}}}
\newrefformat{sec}{\hyperref[#1]{Section~\ref{#1}}}
\newrefformat{subsec}{\hyperref[#1]{Section~\ref{#1}}}
\newrefformat{cha}{\hyperref[#1]{Chapter~\ref{#1}}}
\newrefformat{app}{\hyperref[#1]{Appendix~\ref{#1}}}

\usepackage{MnSymbol}
\DeclareMathAlphabet\mathbb{U}{msb}{m}{n}

\usepackage{booktabs}% professional-quality tables
\usepackage[title]{appendix}

\usepackage[force]{feynmp-auto}

\makeatother

\begin{document}
\renewcommand{\figurename}{FIG.}\renewcommand{\tablename}{TABLE}\renewcommand{\appendixname}{APPENDIX}
\global\long\def\R{\mathbb{R}}%
\global\long\def\llangle{\langle\!\langle}%
\global\long\def\rrangle{\rangle\!\rangle}%
\global\long\def\T{\mathrm{\mathrm{T}}}%
\global\long\def\tran{\mathrm{^{\mathrm{T}}}}%
\global\long\def\lowerindex{\vphantom{X^{N}}}%
\global\long\def\extralowerindex{\vphantom{X^{N^{N}}}}%
\global\long\def\symm{\operatorname{sym}}%

\title{Learning Interacting Theories from Data}
\author{Claudia Merger}
\email{c.merger@fz-juelich.de}

\affiliation{Institute of Neuroscience and Medicine (INM-6) and Institute for Advanced
Simulation (IAS-6) and JARA-Institute Brain Structure-Function Relationships
(INM-10), Jülich Research Centre, Jülich, Germany}
\affiliation{RWTH Aachen University, Aachen, Germany}
\author{Alexandre Ren\'e}
\affiliation{Institute of Neuroscience and Medicine (INM-6) and Institute for Advanced
Simulation (IAS-6) and JARA-Institute Brain Structure-Function Relationships
(INM-10), Jülich Research Centre, Jülich, Germany}
\affiliation{RWTH Aachen University, Aachen, Germany}
\affiliation{Department of Physics, University of Ottawa, Ottawa, Canada}
\author{Kirsten Fischer}
\affiliation{Institute of Neuroscience and Medicine (INM-6) and Institute for Advanced
Simulation (IAS-6) and JARA-Institute Brain Structure-Function Relationships
(INM-10), Jülich Research Centre, Jülich, Germany}
\affiliation{RWTH Aachen University, Aachen, Germany}
\author{Peter Bouss}
\affiliation{Institute of Neuroscience and Medicine (INM-6) and Institute for Advanced
Simulation (IAS-6) and JARA-Institute Brain Structure-Function Relationships
(INM-10), Jülich Research Centre, Jülich, Germany}
\affiliation{RWTH Aachen University, Aachen, Germany}
\author{Sandra Nestler}
\affiliation{Institute of Neuroscience and Medicine (INM-6) and Institute for Advanced
Simulation (IAS-6) and JARA-Institute Brain Structure-Function Relationships
(INM-10), Jülich Research Centre, Jülich, Germany}
\affiliation{RWTH Aachen University, Aachen, Germany}
\author{David Dahmen}
\affiliation{Institute of Neuroscience and Medicine (INM-6) and Institute for Advanced
Simulation (IAS-6) and JARA-Institute Brain Structure-Function Relationships
(INM-10), Jülich Research Centre, Jülich, Germany}
\author{Carsten Honerkamp}
\affiliation{RWTH Aachen University, Aachen, Germany}
\author{Moritz Helias}
\affiliation{Institute of Neuroscience and Medicine (INM-6) and Institute for Advanced
Simulation (IAS-6) and JARA-Institute Brain Structure-Function Relationships
(INM-10), Jülich Research Centre, Jülich, Germany}
\affiliation{RWTH Aachen University, Aachen, Germany}
\date{\today}
\begin{abstract}

One challenge of physics is to explain how collective properties arise
from microscopic interactions. Indeed, interactions form the building
blocks of almost all physical theories and are described by polynomial
terms in the action. The traditional approach is to derive these terms
from elementary processes and then use the resulting model to make
predictions for the entire system. But what if the underlying processes
are unknown? Can we reverse the approach and learn the microscopic
action by observing the entire system? We use invertible neural networks
(INNs) to first learn the observed data distribution. By the choice
of a suitable nonlinearity for the neuronal activation function, we
are then able to compute the action from the weights of the trained
model; a diagrammatic language expresses the change of the action
from layer to layer. This process uncovers how the network hierarchically
constructs interactions via nonlinear transformations of pairwise
relations. We test this approach on simulated data sets of interacting
theories. The network consistently reproduces a broad class of unimodal
distributions; outside this class, it finds effective theories that
approximate the data statistics up to the third cumulant. We explicitly
show how network depth and data quantity jointly improve the agreement
between the learned and the true model. This work shows how to leverage
the power of machine learning to transparently extract microscopic
models from data.
\end{abstract}
\maketitle

\section{Introduction\label{sec:introduction}}

Models of physical systems are frequently described on the microscopic
scale in terms of interactions between their degrees of freedom. Often
one seeks to understand the collective behavior that arises in the
system as a whole. The interactions can feature symmetries, such as
spatial or temporal translation invariance. Prominent examples of
these theories can be found in statistical physics, high energy physics,
but also in neuroscience. The nature of the interactions is often
derived as an approximation of a more complex theory.

The description of systems on the microscopic scale is key to their
understanding. In the absence of an underlying theory, the inverse
problem has to be solved: one needs to infer the microscopic model
by measurements of the collective states. This is typically a hard
problem. A recent route towards a solution comes from studies \citep{leeScaleinvariantRepresentationMachine2022,cohen_separability_2020,duranthonMaximalRelevanceOptimal2021a,goldt_modeling_2020,refinetti_neural_2022,fischer_decomposing_2022,tubiana_emergence_2017,xiao_synergy_2022}
that explore the link between the learned features of artificial neural
networks and the statistics of the data they were trained on. This
inspection yields insights both into the mechanisms by which artificial
neural networks achieve stellar performance on many tasks and into
the nature of the data. In this study, we make the link between learned
parameters and data statistics explicit by studying generative neural
networks.

Generative models learn the statistics which underlie the data they
are trained on.  As such they must possess an internal, learned model
of data which is encoded in the network parameters. In this work,
we gain insights into the nature of the training data by extracting
the model from the network parameters, thus bridging the gap between
the learned model and its interpretation.

One class of generative models are invertible neural networks (INNs),
also called normalizing flows. INNs are invertible mappings trained
to approximate the unknown probability distribution of the training
set \citep{DInh15_1410,dinhDensityEstimationUsing2017}. They can
be used to generate new samples from the same distribution as the
training set, or to manipulate existing data consistent with the
features of the training set (for example, transitions between images
\citep{ardizzoneAnalyzingInverseProblems2019a,odstrcilLINESLogProbabilityEstimation2022,kingmaGlowGenerativeFlow2018a}).
This is achieved by mapping the highly structured input data to a
completely unstructured latent space. The model learned by the network
is expressed through the inverse mapping, as this must generate all
interactions in the data. However, the network mapping is typically
high-dimensional and depends on many parameters, which does not allow
for a direct interpretation.

In this work, we derive interpretable microscopic theories from trained
INNs. We extract an explicit data distribution, formulated in terms
of interactions, from the trained network parameters. These interactions
form the building blocks of the microscopic theory that describes
the distribution of the training data. Furthermore, the process of
extracting the microscopic theory makes the relation between the trained
network parameters and the learned theory explicit. We show how interactions
are hierarchically built through the composition of the network layers.
This approach provides an interpretable relation between the network
parameters and the learned model.

We illustrate and test this framework on several examples where
the underlying theory is exactly known. We find that the networks
are able to learn nontrivial interacting theories. Furthermore, we
show that theories with higher-order interactions emerge as the network
depth increases. Thus, we show how to leverage the power of machine
learning to extract interacting models from data.

Solving inverse problems is a well-known challenge, to which several
approaches have been developed. Previous approaches either rely on
prior knowledge on dynamical systems or stochastic processes such
as specifics of basis functions \citep{jordanEvolvingInterpretablePlasticity2021,confavreuxMetalearningApproachRe2020,bruntonDiscoveringGoverningEquations2016,casadiegoModelfreeInferenceDirect2017,millerLearningTheoryInferring2020}
or the form of update equations \citep{opper_variational_2007,opper_variational_2019,vrettas_variational_2015,ruttor_efficient_2009},
or they are restricted to pairwise couplings between discrete variables
\citep{ackleyLearningAlgorithmBoltzmann1985,mezardExactMeanField2011,schneidmanWeakPairwiseCorrelations2006,zengNetworkInferenceUsing2011,sakellariouEffectCouplingAsymmetry2012,Roudi11,baldassiInverseProblemsLearning2018,decelleInverseProblemsStructured2021,coccoHighdimensionalInferenceGeneralized2011,braunsteinNetworkReconstructionInfection2019},
or they require particular symmetries and specific approximation schemes
to infer higher order couplings \citep{zacheExtractingFieldTheory2020,mastromatteoInverseIsingModel2012}.
In contrast our approach considers continuous variables, is not restricted
to pairwise interactions, and does not require prior knowledge of
the interaction structure.

This paper is structured as follows: in \prettyref{sec:Actions-in-physics}
we introduce the action as the central object of a theory. We then
describe how to extract the action from a trained INN in \prettyref{sec:Learning-Actions-with-INNs}.
Subsequently, we test this framework in several settings in \prettyref{sec:Experiments}.Finally,
we summarize and discuss the main findings, compare our work to different
previously proposed inference schemes, and provide an outlook on how
to extend the framework in \prettyref{sec:discussion}.

\begin{figure}
\centering{}\includegraphics{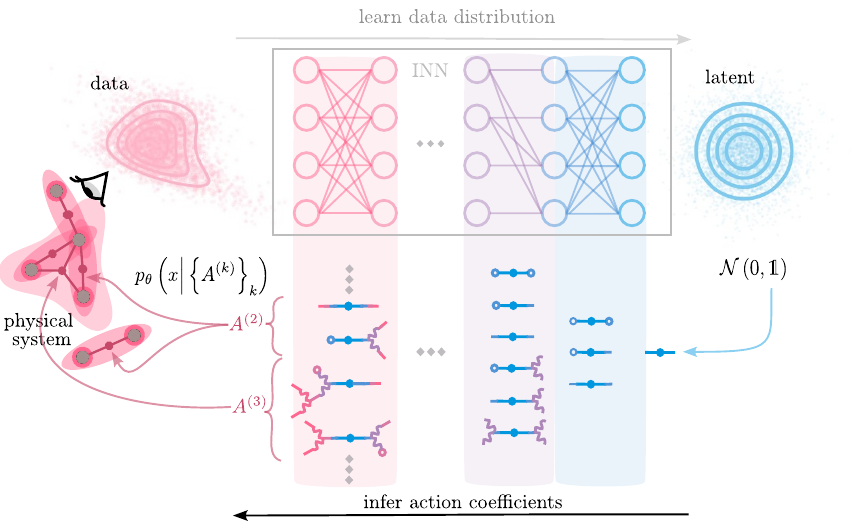}\caption{\textbf{Learning actions from data. }We observe a physical system
of interacting degrees of freedom (gray dots), whose precise interactions
are unknown (shaded areas). We train a neural network on measurements
of the system. The network learns in unsupervised fashion an estimate
of the distribution of training data. We extract the action from the
network parameters layer by layer, using a diagrammatic language.
The final action coefficients $A^{(k)}$ represent the learned interactions
(pink nodes). \label{fig:Schematic-representation}}
\end{figure}

\section{Actions in Physics\label{sec:Actions-in-physics}}

Physical theories are often formulated in terms of microscopic interactions
of many constituents, the degrees of freedom $\left\{ x_{i}\right\} _{1\leq i\leq d}$,
where $d$ is the number of constituents and $x$ describes the system's
state. The degrees of freedom $\left\{ x_{i}\right\} _{1\leq i\leq d}$
can be Ising spins, firing rates of neurons, social agents, or field
points. The interactions of the constituents provide a mechanistic
understanding of the system. The energy or Hamiltonian $H$ of the
system can be written as the sum of these interactions. One can then
ask how probable it is to observe the system in a specific microscopic
state $x$ given an average energy $\langle H\rangle$. The most unbiased
(maximum entropy) estimate of the probability density is then
\[
p_{X}(x)=\frac{1}{Z}e^{-\beta H(x)}\,,
\]
with $Z$ the normalization factor or partition function. Here $p_{X}$
is also known as the Boltzmann distribution \citep{Goldenfeld92}.
In statistical physics, the prefactor $\beta$ is identified as the
inverse temperature.

The action $S_{X}$ of this system is defined as the log probability,
hence $S_{X}(x)=\ln\,p_{X}(x)=-\beta H(x)-\text{\ensuremath{\ln}}Z$.
Therefore the system is fully characterized by $S_{X}$ and measurements
of the system state $x$ correspond to drawing samples from $p_{X}$.
Furthermore, up to the constant prefactor $-\beta$ and the constant
$\text{\ensuremath{\ln}}Z$, the terms in the action are the same
interaction terms as those in the Hamiltonian. 

Consider an action of the form
\begin{align}
S_{X}(x)= & \,A^{(0)}+\sum_{i=1}^{d}A_{i}^{(1)}\,x_{i}+\sum_{i,j=1}^{d}A_{ij}^{(2)}\,x_{i}x_{j}\nonumber \\
 & +\sum_{i,j,k=1}^{d}A_{ijk}^{(3)}\,x_{i}x_{j}x_{k}+\dots\,,\label{eq:poly_action}
\end{align}
where the coefficients $A^{(k)}$ are tensors of rank $k$ and dimension
$d$. Without loss of generality we choose the $A^{(k)}$ to be symmetric
tensors, $A_{i_{1}\dots i_{k}}^{(k)}=A_{\mathcal{P}(i_{1},\dots,i_{k})}^{(k)}$
for any permutation $\mathcal{P}$ of the indices. These coefficients
encode the coordination between the different degrees of freedom
$x_{i}$. In general, we refer to a term of type $A_{i_{1},\dots,i_{k}}^{(k)}\,x_{i_{1}}\cdots x_{i_{k}}$,
as a $k$-point interaction, since this term describes a coaction
of $k$ degrees of freedom for $i_{1},\dots,i_{k}\in\left\{ 1,\dots,d\right\} $
all unequal. In this work, we focus exclusively on actions of the
form of \prettyref{eq:poly_action}, which are suitable only for describing
classical fields $x_{i}$, as the fields and the action coefficients
are tensors and scalars, not operators. However, we are not limited
to equilibrium statistical mechanics: the samples could as well stem
from a time-dependent process; in this case, the action describes
the measure on a path, which is allowed to come from a non-equilibrium
system. Furthermore, even for quantum systems where the interactions
are exactly known, a renormalized classical theory is sometimes sought
to effectively describe the influence of quantum fluctuations \citep{wessel_renormalization_nodate,cuccoli_effective_1995}.

\paragraph{Notation}

In the following, we use the notation $u^{\otimes k}$ for the outer
product of $k$ instances of a tensor $u$ 
\[
u^{\otimes k}=\underbrace{u\otimes u\otimes\cdots\otimes u}_{k\,\text{times}}\,,
\]
and $T^{(k)}\cdot(u)^{\otimes l}$ for $l\leq k$ to denote the
contraction of the first $l$ indices of a rank $k$ tensor $A^{(k)}$
with the first index of each tensor $u$:

\[
\left(A^{(k)}\cdot(u)^{\otimes l}\right)_{\beta_{1},\dots\beta_{l},i_{l+1},\dots,i_{k}}=\sum_{i_{1},\dots,i_{l}}A_{i_{1},\dots,i_{k}}^{(k)}u_{i_{1}\beta_{1}}\cdots\,u_{i_{l}\beta_{l}},
\]
with multi-indices $\beta_{1},\dots,\beta_{l}$ whose rank depends
on the rank of $u$. In the special case that $u$ is a vector, the
indices $\beta_{1},\dots,\beta_{l}$ vanish from the expression. If
additionally $k=l$, the result is a scalar. Hence, \prettyref{eq:poly_action}
becomes
\begin{align*}
S_{X}(x)= & \,A^{(0)}+A^{(1)}\cdot x+\\
 & \,A^{(2)}\cdot(x)^{\otimes2}+A^{(3)}\cdot(x)^{\otimes3}+\dots\,.
\end{align*}

We symmetrize a tensor by averaging over the set $\mathcal{P}(\alpha)$
of all permutations of the multi-index $\alpha$:
\[
\left(\symm A^{(k)}\right)_{\alpha}=\left|\mathcal{P}(\alpha)\right|^{-1}\sum_{\pi\in\mathcal{P}(\alpha)}A_{\pi}^{(k)}\,;
\]
this operation does not change the result of polynomial contractions:
$A^{(k)}\cdot(x_{l})^{\otimes k}\equiv\bigl(\symm A^{(k)}\bigr)\cdot(x_{l})^{\otimes k}$.

A typical objective in statistical physics is understanding how the
microscopic interactions $A^{(k)}$ determine the macroscopic properties
of the system. In this work, we take a different approach: Given
samples from a system with unknown microscopic properties, we extract
the interactions. Generative models such as INNs are a powerful tool
to approximate data distributions $p_{X}$ \citep{DInh15_1410,dinhDensityEstimationUsing2017,papamakariosNormalizingFlowsProbabilistic2021b,kingmaGlowGenerativeFlow2018a}.
In the next section, we demonstrate how we can extract the interaction
coefficients $A^{(k)}$ from trained networks.

\section{Learning Actions with Invertible Neural Networks\label{sec:Learning-Actions-with-INNs}}

\begin{figure*}
\centering{}\includegraphics{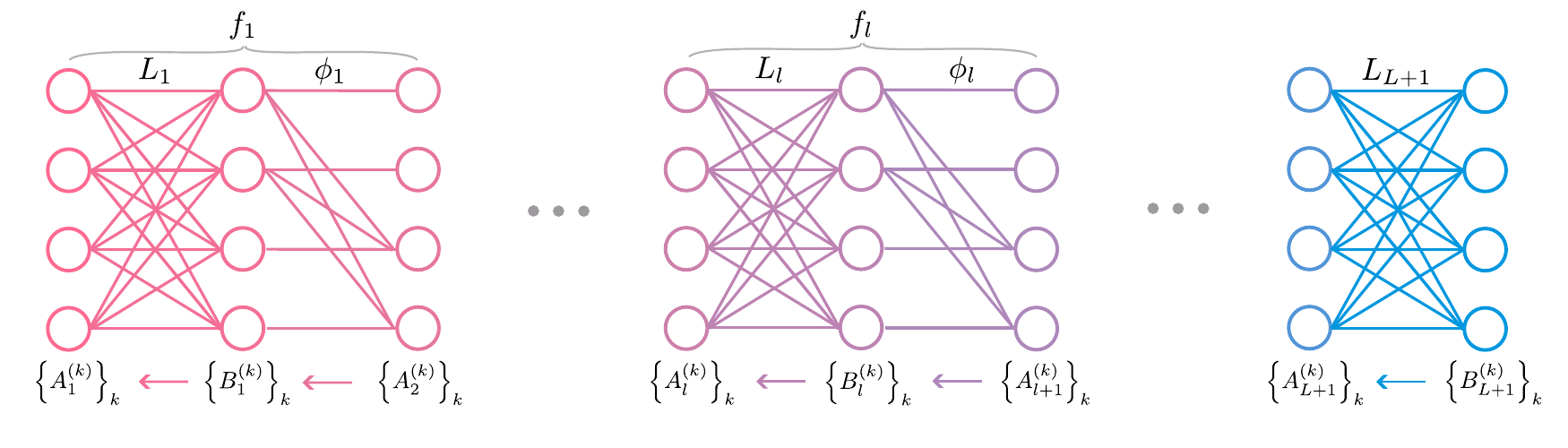}\caption{\textbf{Architecture and coefficient order. }Invertible network composed
of multiple layers. $L_{1},\ldots,L_{L+1}$\textbf{ }denote linear
fully connected layers, $\phi_{1},\ldots,\phi_{L}$\textbf{} are
quadratic nonlinear activation functions. Coefficients $B_{l}^{(k)}$\textbf{
}of the action $S_{H,l}$ of pre--activations $h_{l}$ in layer $l$
of order $k$; coefficients $A_{l}^{(k)}$\textbf{ }of the action
$S_{X,l}$ prior to linear layer \textbf{$l$ }of order $k$.\textbf{\label{fig:Architecture-and-coefficient-order}}}
\end{figure*}

Invertible neural networks are used to learn the data distribution
$p_{X}$ from a data set $\mathcal{D}$ of training samples \citep{DInh15_1410}.
They implement a bijective mapping $f_{\theta}:\mathbb{R}^{d}\rightarrow\mathbb{R}^{d}$,
where the network output $z=f_{\theta}(x)$ is often referred to
as the latent variable. The parameters $\theta$ of the network are
trained such that the latent variables follow a prescribed target
distribution. We follow a common choice for this latent distribution
as a set of $d$ uncorrelated centered Gaussian variables with unit
variance \citep{DInh15_1410}
\begin{equation}
p_{Z}(z)=\exp\left(-\frac{1}{2}z^{\T}z-\frac{d}{2}\ln2\pi\right),\label{eq:output_dist}
\end{equation}
where $z^{\T}z=\sum_{i=1}^{d}z_{i}^{2}$ denotes the Euclidean scalar
product. Given $p_{Z}$, the probability assigned to a specific
input is given by the change of variables formula
\begin{equation}
p_{\theta}(x)=p_{Z}\left(f_{\theta}(x)\right)\left|\det J_{f_{\theta}}\left(x\right)\right|\,,\label{eq:learned_dist}
\end{equation}
which depends only on the network mapping $f_{\theta}$  and its
Jacobian $J_{f_{\theta}}$. The training objective is to minimize
the negative log-likelihood
\begin{align}
\mathcal{L}\left(\theta\right) & =-\sum_{x\in\mathcal{D}}\ln p_{\theta}(x)\nonumber \\
 & =-\sum_{x\in\mathcal{D}}S_{\theta}(x)\,,\label{eq:Loss}
\end{align}
where in the second line we used that $\ln\,p_{\theta}$ is precisely
the action $S_{\text{\ensuremath{\theta}}}$ of the learned distribution.

In this manner, we optimize \prettyref{eq:learned_dist} to approximate
the unknown underlying data distribution using stochastic gradient
descent. Since the target distribution, \prettyref{eq:output_dist}
is a set of independent Gaussians, the mapping $f_{\theta}:x\mapsto z$
of the network aims to eliminate cross-correlations and higher order
dependencies of the components of the latent. On the level of the
action, this means that all $k$-point interactions with $k\geq3$
between components of $z$ must vanish. In turn, the inverse mapping
defines a generative process that induces interactions in the learned
distribution from a non-interacting latent theory.

We now define the architecture that allows us to obtain a polynomial
action $S_{\text{\ensuremath{\theta}}}$ from the network parameters
$\theta$. The network is composed of multiple layers $l$; every
layer mapping $f_{l,\theta}:\mathbb{R}^{d}\rightarrow\mathbb{R}^{d}$
is an invertible function. For each layer, we define an output action
$S_{X,l+1}$, which transforms into the input action $S_{X,l}$ via
the change of variables formula. Using \prettyref{eq:learned_dist}
we compute the input action $S_{X,l}$ of each layer given the output
action $S_{X,l+1}${\small{}
\begin{equation}
S_{X,l}\left(x_{l}\right)=S_{X,l+1}\left(f_{l}\left(x_{l}\right)\right)+\ln\left|\det J_{f_{l}}\left(x_{l}\right)\right|\,,\label{eq:layer_dist_transform}
\end{equation}
}starting with the polynomial action of the latent variable $S_{Z}(y)=\ln p_{Z}(y)$.
We construct all layer mappings $f_{l}$ such that a polynomial action
$S_{X,l+1}$ generates a polynomial action $S_{X,l}$ in \prettyref{eq:layer_dist_transform}
and thus by induction we obtain a polynomial learned input action
$S_{\theta}$.

Each layer mapping $f_{l}$ is composed of a linear mapping $L_{l}$
and a nonlinear mapping\textbf{ $\phi_{l}$}:\textbf{}
\begin{equation}
f_{l}\left(x_{l}\right)=\phi_{l}\circ L_{l}\left(x_{l}\right)\,.
\end{equation}
In the overall architecture, linear and nonlinear mappings are therefore
stacked alternately (see \prettyref{fig:Architecture-and-coefficient-order}).
After the last nonlinear transform $\phi_{L}$, we add an additional
linear transform $L_{L+1}$, such that the network architecture begins
and ends with a linear transform.

The transform of the action via a single layer, \prettyref{eq:layer_dist_transform},
similarly decomposes into two steps, with $h_{l}=L_{l}(x_{l})$ the
intermediate activation: 
\begin{align}
S_{H,l}\left(h_{l}\right) & =S_{X,l+1}\left(\phi_{l}\left(h_{l}\right)\right)+\ln\left|\det J_{\phi_{l}}\left(h_{l}\right)\right|\,,\label{eq:layer_dist_transform_S_H}\\
S_{X,l}\left(x_{l}\right) & =S_{H,l}\left(L_{l}\left(x_{l}\right)\right)+\ln\left|\det J_{L_{l}}\left(x_{l}\right)\right|\,.\label{eq:layer_dist_transform_S_X}
\end{align}
The transformations from $S_{X,l+1}$ to $S_{H,l}$, and from $S_{H,l}$
to $S_{X,l}$ are therefore determined by $\phi_{l}$ and $L_{l}$,
respectively, which we express schematically as
\begin{equation}
S_{X,l+1}\xrightarrow{\phi_{l}}S_{H,l}\overset{L_{l}}{\longrightarrow}S_{X,l}\,.\label{eq:Action_mapping_order}
\end{equation}
The remainder of this section is concerned with expressing \eqref{eq:Action_mapping_order}
in terms of transforms of the action coefficients.

Since the actions are polynomials, at each step we can write\textbf{
$S_{X,l},\,S_{H,l}$} in terms of coefficients  $\left\{ A_{l}^{(k)}\right\} _{k},\left\{ B_{l}^{(k)}\right\} _{k}$:
\begin{align}
S_{X,l}(x_{l}) & =\sum_{k=0}^{K_{l}}A_{l}^{(k)}\cdot(x_{l})^{\otimes k}\,,\label{eq:S_X}\\
S_{H,l}(h_{l}) & =\sum_{k=0}^{K_{l}}B_{l}^{(k)}\cdot(h_{l})^{\otimes k}\,,\label{eq:S_H}
\end{align}
where the sum runs over the ranks $k$ of the coefficient tensors.
Here $K_{l}$ is the degree of the polynomial in $h_{l}$, which depends
on the layer index $l$. We show in the following that the rank of
the polynomial does not change between\textbf{ }$S_{X,l}$ and $S_{H,l}$
in the latter step of \prettyref{eq:Action_mapping_order}. The coefficients
further uniquely determine the action. Therefore \prettyref{eq:Action_mapping_order}
is equivalent to the coefficient mapping
\begin{equation}
\left\{ A_{l+1}^{(k)}\right\} _{k}\overset{\phi_{l}}{\longrightarrow}\left\{ B_{l}^{(k)}\right\} _{k}\overset{L_{l}}{\longrightarrow}\left\{ A_{l}^{(k)}\right\} _{k}\,.\label{eq:Coefficient_mapping_order}
\end{equation}
In \prettyref{fig:Architecture-and-coefficient-order} we illustrate
the order in which the coefficients are transformed.

We now derive the recursive equations for the coefficient transforms,
beginning with the linear mapping.

\paragraph{Linear mapping }

The linear mapping is given by
\begin{align}
h_{l}=L_{l}\left(x_{l}\right) & =W_{l}x_{l}+b_{l}\,,\label{eq:linear_transform}
\end{align}
with $W_{l},b_{l}\in\theta$. Combining \prettyref{eq:layer_dist_transform_S_X}
and \prettyref{eq:S_H} yields
\begin{align}
S_{X,l}(x_{l}) & =\sum_{k}B_{l}^{(k)}\cdot(W_{l}x_{l}+b_{l})^{\otimes k}+\ln\left|\det W_{l}\right|\,.\label{eq:action_update_linear}
\end{align}
We find the transformed coefficients $A_{l}^{(k)}$ by expanding
\prettyref{eq:action_update_linear} and ordering them by powers in
$x_{l}$. Higher-rank coefficients of $S_{H,l}$ contribute to lower-rank
coefficients of $S_{X,l}$ via the contraction with the bias $b_{l}$.
All remaining indices of $B_{l}^{(k)}$ must then be contracted with
the first index of $W_{l}$:
\begin{align}
A_{l}^{(k)}= & \left[\sum_{k^{\prime}=0}^{K_{l}-k}\,\binom{k+k^{\prime}}{k^{\prime}}\,B_{l}^{(k+k^{\prime})}\cdot(b_{l})^{\otimes k^{\prime}}\right]\cdot(W_{l})^{\otimes k}\,\nonumber \\
 & +\delta_{k0}\ln|\det W_{l}|\,.\label{eq:coefficient_update_linear}
\end{align}
The combinatorial factor $\binom{k+k^{\prime}}{k^{\prime}}$ arises
due to the symmetry of the coefficients $B^{(k)}$: since they are
symmetric under permutations of the indices, we only need to fix the
number of contractions $k^{\prime}$ with the bias term $b_{l}$ and
count all $\binom{k+k^{\prime}}{k^{\prime}}$ possible combinations
of $k^{\prime}$ indices in $B^{(k)}$. Since $L_{l}$ is linear,
$S_{X,l}$ and $S_{H,l}$ both have the same rank $K_{l}$.

We illustrate \prettyref{eq:coefficient_update_linear} for the final
linear mapping of the network, the first coefficient transform that
starts on the known latent space coefficients $\left\{ B_{L+1}^{(k)}\right\} _{k}$
(on the far right in \prettyref{fig:Architecture-and-coefficient-order}).
The action of the latent distribution given in \prettyref{eq:output_dist}
describes a centered Gaussian with unit covariance; its first three
coefficients are therefore \textbf{$B_{L+1}^{(0)}=-\frac{d}{2}\ln2\pi$},
$B_{L+1}^{(1)}=0$ and $B_{L+1}^{(2)}=-\frac{1}{2}\mathds{1}$, and
all coefficients with rank $k\ge3$ being zero.

The transformed zeroth-rank coefficient $A_{L+1}^{(0)}$, which ensures
that the action stays normalized, reads
\begin{align}
A_{L+1}^{(0)} & =-\frac{d}{2}\ln2\pi+B_{L+1}^{(2)}\cdot(b_{L+1})^{\otimes k}+\ln\left|\det W_{L+1}\right|\nonumber \\
 & =-\frac{d}{2}\ln2\pi-\frac{|b_{L+1}|^{2}}{2}+\ln\left|\det W_{L+1}\right|\,.\label{eq:linear_mapping_example_A0}
\end{align}
The bias in the linear mapping shifts the mean of the probability
distribution, which is induced by the first-order coefficient
\begin{align}
A_{L+1}^{(1)} & =\left[\,\binom{2}{1}\,B_{L+1}^{(2)}\cdot(b_{L+1})^{\otimes1}\right]\cdot(W_{L+1})^{\otimes1}\,\nonumber \\
 & =-W_{L+1}^{\T}b_{L+1}\,.\label{eq:linear_mapping_example_A1}
\end{align}
The second-rank coefficient is transformed in accordance with the
rotation and scaling of the space due to $W_{L+1}$:
\begin{align}
A_{L+1}^{(2)} & =-B_{L+1}^{(2)}\cdot(W_{L+1})^{\otimes2}\,\nonumber \\
 & =-\frac{1}{2}W_{L+1}^{\T}W_{L+1}\,.\label{eq:linear_mapping_example_A2}
\end{align}

In the case of actions with higher-order interactions $k\ge3$, the
coefficient transform \prettyref{eq:coefficient_update_linear} requires
the computation of many more terms. To simplify the coefficient transforms,
we therefore employ a diagrammatic language. A common practice in
statistical physics is to represent $k$-th order interactions as
vertices with $k$ legs \citep{Goldenfeld92}. Accordingly, we here
express tensors of rank $k$ as vertices with $k$ legs and the contraction
between tensors by an attachment of legs. This facilitates the computation
of combinatorial factors, which can be read off from the diagram topology.
The diagrammatic representations of Eqs.~\eqref{eq:linear_mapping_example_A0},
\eqref{eq:linear_mapping_example_A1}, and \eqref{eq:linear_mapping_example_A2}
read:

\begin{align}
A^{(0)}_{L+1}&=
\vcenter{\hbox{\includegraphics[scale=1]{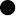}}} 
+ \vcenter{\hbox{\includegraphics[scale=1]{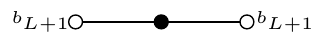}}} 
+\ln\left|\det W_{L+1}\right|   \\
A^{(1)}_{L+1} &=    2\vcenter{\hbox{\includegraphics[scale=1]{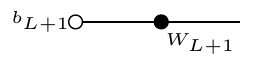}}}  \\
A^{(2)}_{L+1} &= \vcenter{\hbox{\includegraphics[scale=1]{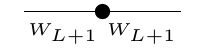}}}
\end{align}A complete presentation of the diagrammatic method is provided in
\prettyref{app:diagrams}.

\paragraph{Nonlinear mapping}

We follow \citet{DInh15_1410} to define an invertible nonlinear
mapping: we split the activation vectors and activation functions
into two halves,\footnote{If the dimension $d$ is uneven, we take the first $\left\lceil \frac{d}{2}\right\rceil $
entries of $h_{l}$ to be in $h_{l}^{1}.$} denoting them by $h_{l}=\begin{pmatrix}h_{l}^{1}\\
h_{l}^{2}
\end{pmatrix}$ and $\phi_{l}=\begin{pmatrix}\phi_{l}^{1}\\
\phi_{l}^{2}
\end{pmatrix}$. The first half is passed onto the next layer unchanged; we then
add a nonlinear function $\tilde{\phi}\left(h_{l}^{1}\right)$ of
the first half onto the second half 
\begin{equation}
x_{l+1}=\phi_{l}(h_{l})=\begin{pmatrix}\phi_{l}^{1}(h_{l})\\
\phi_{l}^{2}(h_{l})
\end{pmatrix}=\begin{pmatrix}h_{l}^{1}\\
h_{l}^{2}
\end{pmatrix}+\begin{pmatrix}0\\
\tilde{\phi}_{l}(h_{l}^{1})
\end{pmatrix}\,.\label{eq:additive_nonlinear_mapping}
\end{equation}
We choose a quadratic nonlinearity, 
\begin{equation}
\tilde{\phi}_{l}(h_{l}^{1})=\tilde{\chi}_{l}\cdot(h_{l}^{1})^{\otimes2}\,,\label{eq:quadratic_nonlinear_mapping}
\end{equation}
where $\tilde{\chi}_{l}\in\mathbb{R}^{\left\lfloor \frac{d}{2}\right\rfloor \times\left\lceil \frac{d}{2}\right\rceil \times\left\lceil \frac{d}{2}\right\rceil }$
is a third-order tensor whose coefficients are trained, $\tilde{\chi}_{l}\in\theta$.
In the following, we will use a shorthand notation $\phi_{l}(h_{l})=h_{l}+\chi_{l}\cdot(h_{l})^{\otimes2}$
with $\tilde{\chi}_{l}$ being the non-zero part of $\chi_{l}$.

Equation~\eqref{eq:quadratic_nonlinear_mapping} is the most elementary
nonlinearity which is compatible with a polynomial action. Through
the composition of $L$ layer transforms $f_{l}$, the network mapping
becomes a polynomial of order $2^{L+1}$. The advantage of decomposing
such a transform in terms of multiple applications of Eqs.~\eqref{eq:linear_transform}
and \eqref{eq:additive_nonlinear_mapping} is a particularly simple
form of the update equations for the coefficients.

Splitting the nonlinear mapping \eqref{eq:additive_nonlinear_mapping}
makes it trivially invertible
\begin{equation}
h_{l}=\phi_{l}^{-1}(x_{l+1})=\begin{pmatrix}x_{l+1}^{1}\\
x_{l+1}^{2}
\end{pmatrix}-\begin{pmatrix}0\\
\tilde{\phi}_{l}(x_{l+1}^{1})
\end{pmatrix}\,,\label{eq:nonlinear_transform_inverse}
\end{equation}
as one can observe by evaluating the composition of Eqs.~\eqref{eq:additive_nonlinear_mapping}
and \eqref{eq:nonlinear_transform_inverse} on an arbitrary vector
$h_{l}$.

We compute the action $S_{H,l}$ from $S_{X,l+1}$ using \prettyref{eq:layer_dist_transform_S_X}.
Since the Jacobian $J_{\phi,l}$ of $\phi_{l}$ is a triangular matrix
with ones on the diagonal, we have $\ln\left|\det J_{\phi,l}\right|=\text{0}.$
Therefore, the transform of the action induced by $\phi_{l}$ is just
the composition
\begin{align}
S_{H,l}(h_{l}) & =S_{X,l+1}\left(h_{l}+\chi_{l}\cdot\left(h_{l}\right)^{\otimes2}\right)\nonumber \\
 & =\sum_{k}A_{l+1}^{(k)}\cdot\left(h_{l}+\chi_{l}\cdot\left(h_{l}\right)^{\otimes2}\right)^{\otimes k}\,.\label{eq:action_update_nonlinear}
\end{align}
 Equation \eqref{eq:action_update_nonlinear} yields a polynomial
of order $K_{l}=2K_{l+1}$. We expand the products in \prettyref{eq:action_update_nonlinear}
and reorder the terms to obtain the transform of the action coefficients.
Since each factor of $\chi_{l}$ increases the rank of the resulting
tensor by one, lower-order coefficients $A_{l+1}^{(k-k^{\prime})}$
contribute to the coefficient $B_{l}^{(k)}$ via
\[
\binom{k-k^{\prime}}{k^{\prime}}\,A_{l+1}^{(k-k^{\prime})}\cdot(\chi_{l})^{\otimes k^{\prime}}\,,
\]
with $k>k^{\prime}\geq1$. Each contraction of $A_{l+1}^{(k-k^{\prime})}$
with $\chi_{l}$ consumes one index in $A_{l+1}^{(k-k^{\prime})}$
and the first index in $\chi_{l}$, but adds two indices to the resulting
tensor. As a result, for each $\chi_{l}$ in the contraction, the
rank is raised by one. Therefore $k^{\prime}$ factors of $\chi_{l}$
are needed to increase the rank from $k-k^{\prime}$ to $k$. The
factor $\binom{k-k^{\prime}}{k^{\prime}}$ arises because there are
$\binom{k-k^{\prime}}{k^{\prime}}$ ways of choosing $k^{\prime}$
of the $k-k^{\prime}$ indices of the tensor to which to contract
the factors of $\chi_{l}$. However, contractions of this type are
no longer symmetric tensors because the resulting $k$-th order tensor
has $2k^{\prime}$ indices stemming from $\chi_{l}$ and the remaining
ones from $A^{(l+1,k-k^{\prime})}$. To express this result as a symmetric
tensor, we symmetrize the result. This yields
\begin{align}
B_{l}^{(k\leq1)} & =A_{l+1}^{(k)}\nonumber \\
B_{l}^{(k>1)} & =\symm\sum_{k^{\prime}=0}^{k}\,\binom{k-k^{\prime}}{k^{\prime}}\,A_{l+1}^{(k-k^{\prime})}\cdot(\chi_{l})^{\otimes k^{\prime}}\,.\label{eq:coefficient_update_nonlinear}
\end{align}

Diagramatically, the contraction with $\chi_{l}$ is represented by
splitting the legs of a vertex. By counting the number of splits we
can therefore infer the number of factors $\chi_{l}$ of any diagram.
To illustrate this, we here show the mapping $\left\{ A_{L+1}^{(k)}\right\} _{k}\overset{\phi_{L}}{\longrightarrow}\left\{ B_{L}^{(k)}\right\} _{k}$,
which generates interactions up to the fourth order. The zeroth- and
first-rank coefficients remain unchanged $B_{L}^{(0)}=A_{L+1}^{(0)}$
and $B_{L}^{(1)}=A_{L+1}^{(1)}$, as the $A_{L+1}^{(0)}$ has no legs
to split; the splitting of legs in $A_{L+1}^{(1)}$ gives a contribution
to $B_{L}^{(2)}$:\begin{align}
B^{(2)}_L  =  & \vcenter{\hbox{\includegraphics[scale=1]{Figures/Diagrams/A_2_extended_legs.pdf}}}  \nonumber \\
&+ 2 \vcenter{\hbox{\includegraphics[scale=1]{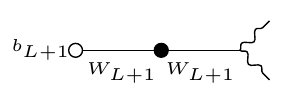}}}
\end{align}This diagrammatic expression corresponds to
\begin{align*}
B_{L}^{(2)}= & B_{L+1}^{(2)}\cdot\left(W_{L+1}\right)^{\otimes2}\\
 & +\symm\left(\left[B_{L+1}^{(2)}\cdot b_{L+1}\right]\cdot W_{L+1}\right)\cdot\chi_{L}
\end{align*}
No further diagrams are generated from $A_{L+1}^{(1)}$, as all legs
are split. The higher-order interactions $B_{L}^{(3)},\,B_{L}^{(4)}$
emerge through the splitting of legs in $A_{L+1}^{(2)}$\begin{align}
B^{(3)}_L  = & \,\binom{2}{1}\vcenter{\hbox{\includegraphics[scale=1]{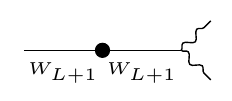}}}\, ,\\
B^{(4)}_L  = & \,\vcenter{\hbox{\includegraphics[scale=1]{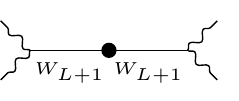}}}
\end{align}In tensor notation, the same expressions read
\begin{align*}
B_{L}^{(3)} & =\,\binom{2}{1}\,\symm\left[B_{L+1}^{(3)}\cdot\left(W_{L+1}\right)^{\otimes2}\right]\cdot\chi_{L}\,,\\
B_{L}^{(4)} & =\,\symm\left[B_{L+1}^{(3)}\cdot\left(W_{L+1}\right)^{\otimes2}\right]\cdot\left(\chi_{L}\right)^{\otimes2}\,.
\end{align*}
This exemplifies how the interactions are built hierarchically as
further layer transforms contribute contractions with $W_{l\leq L}$,
$b_{l\leq L}$ and $\chi_{l<L}$, to previous coefficients. See \prettyref{app:diagrams}
for further details.

The degree $K_{l}$ of the action doubles with each layer, starting
from the output action with degree two. Networks of depth $L$ thus
generate actions of degree $2^{L+1}$. Through the composition of
several nonlinear mappings like \prettyref{eq:additive_nonlinear_mapping},
the prefactors $\chi_{l}$ of later layers will be exponentiated
alongside their activations. Increasing the number of layers will
therefore generate terms of arbitrarily high degree in both $x$ and
in the prefactors. For example, the contribution of $\chi_{1}$ to
$x_{L}$ will be of order $(\chi_{1})^{2^{L-1}}$. As a result, large
values in $\chi_{l}$ are unfavorable as they make the activations
diverge. In practice, we find that the entries of the tensors $\chi_{l}$
of trained networks are typically small, $\left|\left(\chi_{l}\right)_{ijk}\right|\ll1$.

We note that all coefficients with rank $k>2$ must contain at least
$k-2$ factors of $\chi_{l}$, where the different factors in general
originate from different layers. These terms of rank $k>2$ constitute
the non-Gaussian part of the action. Therefore, for applications
where the data can be described as a perturbed Gaussian, small entries
in $\chi_{l}$ are sufficient. The higher the rank of the coefficient,
the smaller its entries. Consequently, we place a cutoff of two on
the number of factors of $\chi_{l}$ in the action coefficients, thus
ignoring negligible contributions. This effectively imposes a maximum
rank of $k=4$ onto the action coefficients.

Coefficients of high rank can be numerically intractable for large
dimension $d$, as their size grows as $\mathcal{O}\left(d^{k}\right)$.
To mitigate this, we make use of \prettyref{eq:quadratic_nonlinear_mapping}
to write the coefficients in a decomposed form, focusing on the most
significant contributions, which speeds up the computations and allows
for tractable reductions in the size of the stored tensors. We specify
this decomposition in \prettyref{app:decomposed_tensors}.

We began this section by equating the transform of the action through
the network to the transform of its coefficients, decomposed as alternating
linear and nonlinear transforms. We then made these transforms explicit
in Eqs.~\eqref{eq:coefficient_update_linear} and \eqref{eq:coefficient_update_nonlinear}.
Given a trained network, these expressions allow us to extract the
learned action through the iterative application of the coefficient
transforms from the last layer to the first. In this way, we can describe
the data distribution constructively, by tracking how the latent distribution
is transformed through successive network layers.

Equation \eqref{eq:coefficient_update_nonlinear} shows how higher-rank
coefficients hierarchically emerge through repeated contractions with
the parameters $\chi_{L},\dots,\chi_{1}$ of the nonlinear mappings
and $W_{L},\dots,W_{1},b_{L},\dots,b_{1}$ of the linear mappings
of different layers.  In the data space, these coefficients correspond
to interactions; therefore this approach establishes an explicit
relation between network parameters $\theta$ and the characteristic
properties of the learned distribution $p_{\theta}.$ In the next
section, we test this method in several cases with known ground-truth
distributions.

\section{Experiments\label{sec:Experiments}}

In this section, we will test the learning of actions in three different
settings. In \prettyref{subsec:In-class-distributions}, we use a
randomly initialized teacher network to generate samples. The teacher
network has the same architecture as the one described in \prettyref{sec:Learning-Actions-with-INNs},
enabling us to compute the ground truth action. In \prettyref{subsec:Out-of-class-distributions},
the ground truth action coefficients themselves are generated randomly,
leading to multimodal data distributions. Finally in \prettyref{subsec:Interaction-on-a-lattice},
we study a physics-inspired model system with interactions on a square
lattice of $d=10^{2}$ sites.

\subsection{In-class distributions\label{subsec:In-class-distributions}}

\begin{figure}
\centering{}\includegraphics{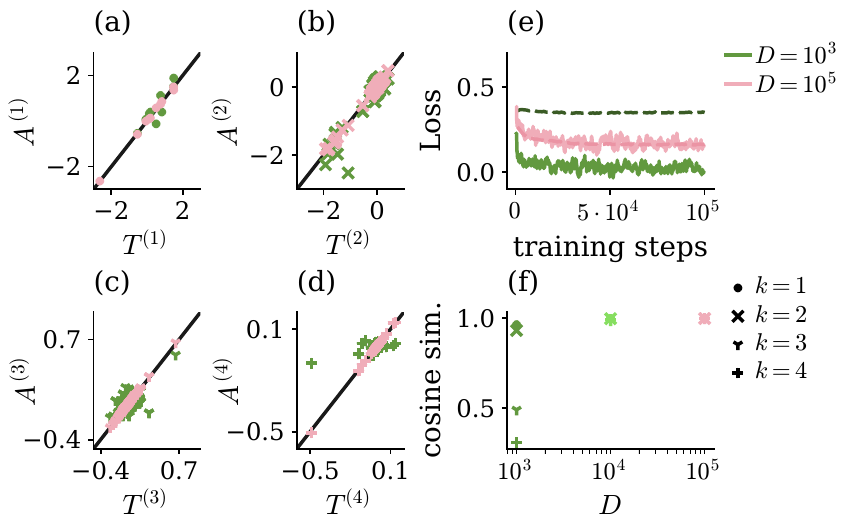}\caption{\textbf{\label{fig:Teacher-Student}Teacher-student coefficient comparison}
for varying training set sizes $D=|\mathcal{D}|$. Both teacher and
student have depth $L=1$.\textbf{ (a--d)} Student coefficients $A^{(k)}$
over teacher coefficients $T^{(k)}$ up to fourth order for $D=10^{3}$
in green and $D=10^{5}$ in pink.\textbf{ (e)} Training loss (full
lines) and test loss (dashed lines) over training steps.\textbf{ (f)}
Cosine similarity of coefficients over number of training samples.}
\end{figure}

 First, we test whether we can recover the action coefficients of
a known action from samples drawn from the respective distribution.
We initialize a teacher network with random weights and compute the
corresponding action coefficients $\left\{ T^{(k)}\right\} _{k\leq4}$
with the method outlined in \prettyref{sec:Learning-Actions-with-INNs}.
We then generate a training set $\mathcal{D}$ by sampling Gaussian
random variables $z\sim p_{Z}$ and apply the inverse network transform
of the teacher on them; the elements of $\mathcal{D}$ are therefore
samples from the teacher distribution. Since the teacher distribution
is by construction part of the set of learnable student distributions,
we refer to this as an in-class distribution. We then initialize
a student network to identity $W_{l}=\mathds{1},\,b_{l}=0,\,\chi_{l}=0\,\forall l$
and train it on the training set $\mathcal{D}$. This choice for the
initialization ensures that all variability in the trained result
is due to the training data and the sequence of random batches drawn
during training. After training, we extract the student coefficients
$\left\{ A^{(k)}\right\} _{k\leq4}$ as described in \prettyref{sec:Learning-Actions-with-INNs}.
The student network has learned the teacher distribution if the associated
action coefficients match, $T^{(k)}=A^{(k)}\,\forall k$.

Note that $T^{(k)}=A^{(k)}\,\forall k$ does not imply that the parameters
$\theta$ of the teacher and student network are equal. Due to the
rotational invariance of the latent space, an additional linear transform
which rotates the latent space $z$ does not result in a change in
the learned action. Hence $T^{(k)}=A^{(k)}\,\forall k$ implies only
that the two networks learn the same statistics.

We compare teacher and student coefficients in \prettyref{fig:Teacher-Student}
for two different training set sizes $\left|\mathcal{D}\right|$.
For a sufficiently large data set, the student can learn the teacher
coefficients arbitrarily well: In \prettyref{fig:Teacher-Student}~(a--d),
the coefficient entries coincide while the network parameters $\theta$
do not align (see \prettyref{app:Dissimilarity-of-parameters} for
a comparison between network parameters). This confirms that the extracted
coefficients are indeed characteristic of what the network has learned,
as opposed to the parameters. Given sufficient samples, we therefore
find that the method recovers the correct action coefficients.

For a smaller training set with $\lvert\mathcal{D}\rvert=10^{3}$,
we find that the student network overfits the training set. To see
this, we compute the test loss $\mathcal{L}_{\mathrm{test}}(\theta)$
on a test set of $10^{4}$ samples. In \prettyref{fig:Teacher-Student}~(e)
the test loss is significantly larger than the training loss. This
is reflected in deviating coefficient entries in \prettyref{fig:Teacher-Student}~(a--d).
To quantify this disparity, we compute the cosine similarity between
the tensors
\begin{equation}
\text{\ensuremath{\cos\angle}}\left(T^{(k)},S^{(k)}\right)=\frac{\left|\sum_{\alpha}T_{\alpha}^{(k)}S_{\alpha}^{(k)}\right|}{\sqrt{\sum_{\alpha}\left(S_{\alpha}^{(k)}\right)^{2}\sum_{\alpha}\left(T_{\alpha}^{(k)}\right)^{2}}}\label{eq:cosine_sim_def}
\end{equation}
where the sum runs over all independent indices $\alpha$, excluding
duplicate tensors entries which are equal due to the symmetry. The
cosine similarity ranges between zero and one, for perfect alignment
it is equal to one. \prettyref{fig:Teacher-Student}~(f) shows how
the cosine similarity between teacher and student coefficients increases
with the training set size. The lower order coefficients $T^{(k\leq2)}$
are approximated well even in the case of little training data. The
learned higher-order coefficients clearly deviate from the teacher
coefficients in this case (see \prettyref{fig:Teacher-Student}~(c,d,f)),
indicating that the higher-order coefficients, corresponding to higher-order
interactions in the teacher distribution, can only be conveyed through
larger data sets.

\paragraph{Learning rules in coefficient space}

 Higher-order interactions depend on higher-order statistics of the
data distribution, which must be expressed through a limited amount
of samples. We train the network using stochastic gradient descent
(SGD), which updates all parameters of the network at training time
$t$ according to the dependence of the loss $\mathcal{L}$ on a subset
of training data $\mathcal{D}_{t}\subset\mathcal{D}.$ In SGD, the
update of a single weight $\Delta\theta_{i}=\theta_{i}(t+1)-\theta_{i}(t)$
is
\begin{align}
\Delta\theta_{i} & =-\eta\frac{\partial}{\partial\theta_{i}}\mathcal{L}_{\mathcal{D}_{t}}\nonumber \\
 & =\eta\frac{\partial}{\partial\theta_{i}}\langle S_{\theta}(x)\rangle_{\mathcal{D}_{t}},\label{eq:parameter_updates}
\end{align}
where $\eta$ is the learning rate and $\langle\cdot\rangle_{\mathcal{D}_{t}}$
denotes the average over the current training batch $\mathcal{D}_{t}$.
In \prettyref{app:Coefficient-update-equation}, we show that this
leads to a noisy update in the coefficients $A^{(k)}$ for $k\geq1$
of 
\begin{widetext}
\begin{align}
\Delta A_{\alpha}^{(k)}= & \eta\left(\xi_{\alpha,t}^{(k)}+\bigl\langle\left(x^{\otimes k}\right)_{\alpha}\bigr\rangle_{\mathcal{D}}-\bigl\langle\left(x^{\otimes k}\right)_{\alpha}\bigr\rangle_{A}\right)\sum_{i}\left(\frac{\partial A_{\alpha}^{(k)}}{\partial\theta_{i}}\right)^{2}\nonumber \\
&  + \eta \sum_{l,\alpha_l \neq \alpha} \left(\xi_{\alpha_l,t}^{(l)}+\bigl\langle\left(x^{\otimes l}\right)_{\alpha_l}\bigr\rangle_{\mathcal{D}}-\bigl\langle\left(x^{\otimes l}\right)_{\alpha_l}\bigr\rangle_{A}\right) \sum_{i} \frac{\partial A_{\alpha}^{(k)}}{\partial\theta_{i}}\frac{\partial A_{\alpha_l}^{(l)}}{\partial\theta_{i}}  +\mathcal{O}(\Delta\theta^{2})\,,\label{eq:noisy_update_coefficients}
\end{align}
\end{widetext}
where $\langle\cdot\rangle_{\mathcal{D}}$ is the empirical average
over all samples in the full training set $\mathcal{D}$, and $\langle\cdot\rangle_{A}$
is the expectation with regard to the current estimate of the density
depending on learned coefficients $\left\{ A^{(k)}\right\} _{k}$.
The random variable $\xi_{t}^{(k)}$ encodes the difference between
the mean estimated on the whole training set and a training batch.
One can show that on average over all batches, the noise vanishes,
$\langle\xi^{(k)}\rangle=0$ , and the variance also decreases with
the training set size
\[
\bigl\langle\!\bigl\langle\xi^{(k)}\bigr\rangle\!\bigr\rangle=\left(\bigl\langle x^{\otimes 2k}\bigr\rangle_{\mathcal{D}}-\left(\bigl\langle x^{\otimes k}\bigr\rangle_{\mathcal{D}}\right)^{\otimes2}\right)|\mathcal{D}_{t}|^{-1}
\]
 (see e.g. \citep{cramerMathematicalMethodsStatistics1999}). Smaller
batch sizes $|\mathcal{D}_{t}|$ therefore increase the noise in the
updates of the action coefficients.

The expected update $\bigl\langle\text{\ensuremath{\Delta A_{\theta}^{(k)}\bigr\rangle}}$
vanishes on average over all batches when the learned moments and
the moments on the training set match: $\langle x^{\otimes k}\rangle_{\mathcal{D}}=\langle x^{\otimes k}\rangle_{A}$.
However, for any finite training set, there will furthermore be a
deviation between $\langle x{}^{\otimes k}\rangle_{\mathcal{D}}$
and the true moment $\langle x{}^{\otimes k}\rangle_{T}$ of the teacher
network. This expected difference scales as 
\[
\langle x{}^{\otimes k}\rangle_{\mathcal{D}}-\langle x{}^{\otimes k}\rangle_{T}\propto\sqrt{\left[\langle x{}^{\otimes2k}\rangle_{T}-\left(\langle x{}^{\otimes k}\rangle_{T}\right)^{\otimes2}\right]|\mathcal{D}|^{-1}}\,.
\]
Therefore, not only will there be a batch-size dependent variability
in the coefficient updates, but a bias induced by the limited amount
of training data. This drift introduces a bias in the training, leading
to overfitting. For many distributions, both $\langle x{}^{\otimes k}\rangle_{\mathcal{D}}-\langle x{}^{\otimes k}\rangle_{T}$
and $\langle\langle\xi^{(k)}\rangle\rangle$ increase with $k$ \citep{krasilnikovAnalysisEstimationErrors2019a,beregunApplicationCumulantCoefficients2018}.
In this case, both the bias and the variability of the training procedure
increase with $k$, explaining why it is harder to learn higher-order
statistics.

\subsection{Out-of-class distributions\label{subsec:Out-of-class-distributions}}

\begin{figure*}
\centering{}\includegraphics{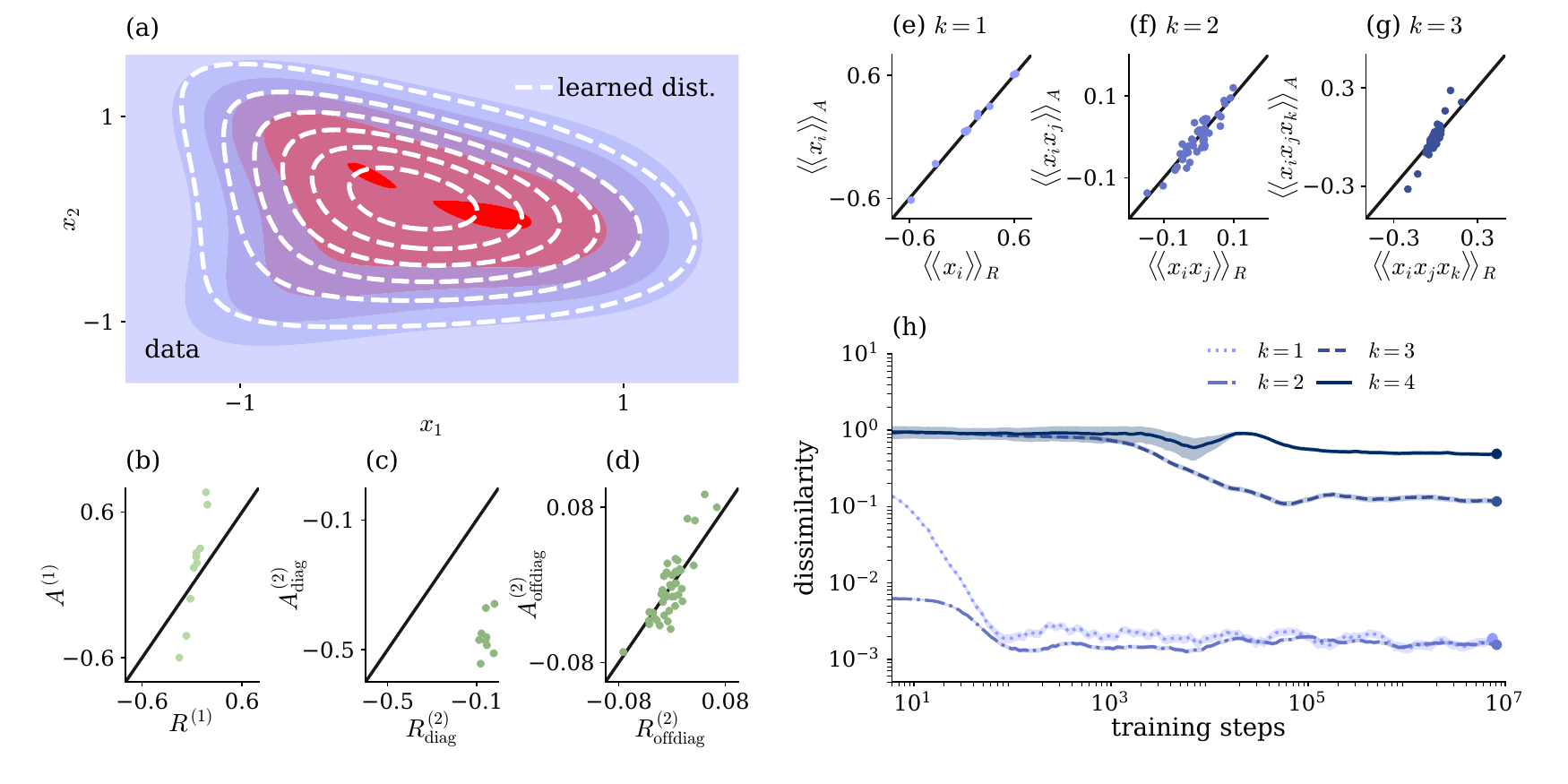}\caption{\textbf{Learning an effective monomodal theory}.\textbf{ (a)} Two-dimensional
example of random density with multiple maxima. White lines are level
lines of learned distribution for a five layer network. All other
panels show results from a $d=10$-dimensional data set.\textbf{ (b--d)}
Learned over true coefficients for a three layer network on a $d=10$.
We distinguish diagonal tensor entries from off-diagonal ones, where
at least two indices differ.\textbf{ (e--g)} Learned over true cumulants,
computed from samples. Error bars are typically smaller than marker
size.\textbf{ (h)} Dissimilarity of true and learned cumulants: $1-\text{\ensuremath{\cos\angle}}\left(\langle\!\langle x^{\otimes k}\rangle\!\rangle_{A},\langle\!\langle x^{\otimes k}\rangle\!\rangle_{R}\right)$
over training steps. We record the cumulants at logarithmically spaced
intervals during training. The curves are then smoothed by averaging
over ten adjacent recording steps. Shaded areas show the variation
due to the estimation of the cumulants from samples. Dots indicate
training stage of cumulants shown in \textbf{(e--g)}.\label{fig:RandomCoeffiicients}}
\end{figure*}

In the previous section, we have demonstrated that invertible networks
accurately learn any distribution generated by the image set of inverse
mappings $f_{\theta}^{-1}$. However, it is interesting to investigate
how our approach deals with distributions outside this set. A first
step to this end lies in understanding the nature of mappings employed
by the invertible network. 

The proposed network architecture belongs to the class of volume-preserving
networks \citep{DInh15_1410,dinhDensityEstimationUsing2017}: the
additive nature of the nonlinearity in \prettyref{eq:additive_nonlinear_mapping}
leads to a constant Jacobian determinant $\det J_{f_{\theta}}(x)$.
The Jacobian determinant $\det J_{f_{\theta}}(x)$ of a mapping states
how the image of the mapping is locally stretched. Since the only
contribution to $\left|\det J_{f_{\theta}}(x)\right|$ comes from
the linear transform, \prettyref{eq:linear_transform}, the Jacobian
determinant is constant. Therefore, this stretch is homogeneous everywhere.
As a result, an invertible network with $\left|\det J_{f_{\theta}}(x)\right|=const.$
and a Gaussian target distribution $p_{Z}$ can only learn unimodal
distributions $p_{\theta}$, i.e. distributions with only one maximum.
To see this, we compute the gradient of \prettyref{eq:learned_dist}
with respect to $x$:
\[
\nabla_{x}p_{\theta}(x)=0\,\overset{\left|\det J_{f_{\theta}}(x)\right|=const.>0}{\Longleftrightarrow}\nabla_{f_{\theta}}p_{Z}\left(f_{\theta}(x)\right)=0\,.
\]
This shows that the learned input distribution $p_{\theta}(x)$ has
an extremum at $x_{0}$ if and only if the target distribution has
an extremum at $f_{\theta}(x_{0})$. Since $p_{Z}$ has a single extremum,
so does $p_{\theta}.$ This limitation is not unique to the choice
of polynomial activations, but a consequence of the special structure
of the Jacobian of the nonlinearity \prettyref{eq:additive_nonlinear_mapping}
and the latent distribution $p_{Z}$.

Since not every distribution is unimodal, we discuss in \prettyref{sec:discussion}
how to extend the framework introduced here to incorporate multimodal
distributions. An effective unimodal model, however, may also prove
useful. While volume-preserving invertible networks with unimodal
latent distribution $p_{Z}$ cannot learn a multimodal distribution
$p_{\theta}$ exactly, they can learn an approximation. In this section
we show how volume preserving networks can therefore be used to extract
an effective theory in the multimodal case.

To avoid tieing results to a particular choice of distribution, we
generate actions $S_{R}$ with random coefficients $\left\{ R^{(k)}\right\} _{k\leq3}$.
A diagonal negative action coefficient $R^{(4)}$ is then added to
obtain a normalizable distribution. We ensure that the corresponding
distributions are multimodal and that their terms are balanced in
strength using a sampling method for the coefficients that is detailed
in \prettyref{app:generating-random-actions}. Although the action
$S_{R}$ is an unnormalized log-probability\footnote{We do not compute the constant term in $S_{R}$ which ensures $\int dx\exp\left(S_{R}(x)\right)=1$
as it is not needed for the sampling method.}, we can then sample a training set $\text{\ensuremath{\mathcal{D}}}$
using Markov chain Monte Carlo; for this work we used a Hamiltonian
Monte Carlo \citep{duaneHybridMonteCarlo1987,nealMCMCUsingHamiltonian2012,betancourtConceptualIntroductionHamiltonian2017}
sampler implemented in PyMC3 \citep{salvatierProbabilisticProgrammingPython2016}.
(For details see Appendix~\ref{app:Sampling-random-actions}.) 

Given a known random action $S_{R}$ and a corresponding training
set $\mathcal{D}$ of generated samples, we train networks of different
depths and compare their action coefficients. In \prettyref{fig:RandomCoeffiicients}\textbf{
(a)} we show a two-dimensional example of such a randomly generated
distribution as well as the learned monomodal approximation. Figure
\prettyref{fig:RandomCoeffiicients}\textbf{ (b)}-\textbf{(d)} show
comparisons of the true compared to the learned coefficients in the
$d=10$ dimensional case. As expected, some action coefficients cannot
be learned correctly. The largest deviations from the true coefficients
occur in the diagonal entries $A_{i}^{(1)},A_{ii}^{(2)},\ldots$.
However, we observe that many action coefficient entries that have
at least two different indices (shown as $A_{\mathrm{offdiag.}}^{(2)}$in
\prettyref{fig:RandomCoeffiicients} \textbf{(d)}) recover approximately
the correct value.

Using the network to generate samples, which hence belong to the
learned distribution $p_{\theta}$, we compare the cumulants of the
two distributions. The cumulants of a distribution can be computed
from its moments and vice versa. For example, the first three cumulants
are equal to the mean, the variance, and the centered third moment
of a distribution. Cumulants are better suited than moments to compare
two different distributions because they contain only independent
statistical information. We distinguish $k$-th order cumulants from
moments by using single brackets $\langle x^{k}\rangle$ for moments
and double brackets $\langle\!\langle x^{k}\rangle\!\rangle$ for
cumulants.

Despite the disparity in the coefficients, we find that the cumulants
agree up to third order (compare panels \textbf{(e)} - \textbf{(g)}
in \prettyref{fig:RandomCoeffiicients}). The learned distribution
is therefore an effective theory that reproduces the statistics of
the system beyond the Gaussian order, since in a Gaussian model the
third-order cumulants are zero. Equation \eqref{eq:noisy_update_coefficients}
shows that the action coefficient $A^{(k)}$ converges in expectation
either when the moments of the training set and the learned distribution
coincide, $\langle x{}^{\otimes k}\rangle_{\mathcal{D}}=\langle x^{\otimes k}\rangle_{A}$,
or when the network cannot tune the coefficients in the relevant direction.
Therefore the training aims to match the moments $\langle x{}^{\otimes k}\rangle_{\mathcal{D}},\langle x^{\otimes k}\rangle_{A}$
(and thereby, the cumulants) within the bounds of the flexibility
allowed by the network architecture. We find that the higher order
cumulants are learned later in training (see \prettyref{fig:RandomCoeffiicients}\textbf{(h)}).

Mulitmodality often appears as a result of symmetry breaking. Consider
the classical example of an Ising model \citep{Amit84}. The action
of this system is symmetric under a global flipping of all spins.
Below the critical temperature, two modes appear, one for positive
and one for negative net magnetization. However, in a physical system,
this multimodality cannot be observed, because the probability of
a global sign flip approaches zero as the system size increases. Furthermore,
external factors such as coupling to the environment or a measurement
device will also break the symmetry. In such a setting, the network
can nevertheless find an informative theory, characterizing the observed
monomodal distribution.

\subsection{Interaction on a lattice\label{subsec:Interaction-on-a-lattice}}

Physical theories often feature a local structure of the interactions,
for example, a lattice structure. We here construct such a system
by introducing nearest-neighbor couplings on a square lattice of $d=10\times10$
sites with periodic boundary conditions. Furthermore, we introduce
self-interaction terms of second and fourth order. The resulting action
is symmetric under a global sign change and under translations along
the lattice. In an experiment, such symmetries may be broken by the
coupling of the system to an external environment. We model this breaking
of both symmetries by introducing a heterogeneous external field which
introduces a bias  to each degree of freedom.

The action therefore reads
\begin{align}
S_{I} & (x)=-\beta\left[\frac{1}{2}\sum_{i,j}x_{i}\left(r_{0}\delta_{ij}-\Lambda_{ij}\right)x_{j}+\sum_{i}\left(h_{i}x_{i}+ux_{i}^{4}\right)\right]\,\nonumber \\
 & =:I^{(1)}\cdot x+I^{(2)}\cdot\left(x\right)^{\otimes2}+I^{(4)}\cdot\left(x\right)^{\otimes4}\,,\label{eq:action_lattice}
\end{align}
where the $I^{(k)}$ are the coefficients of the true distribution
and we have omitted the normalization. Here $\beta$ serves as an
inverse temperature. The matrix Laplacian $\Lambda_{ij}=-\delta_{ij}\text{deg}(i)+a_{ij}$
with $a_{ij}$ the adjacency matrix ($a_{ij}=1$ if $i,j$ are connected
and $a_{ij}=0$ else) constitutes an interaction with the $\deg(i)=4$
nearest neighbors on the lattice. Both the diagonal part of $\Lambda$
and $r_{0}$ encode a second-order self-interaction.

The fourth-order term is another self-interaction $-\beta ux_{i}^{4}$.
 This model can be considered the lattice version of the effective
long distance theory of an Ising model in two dimensions \citep{Amit84}.
We illustrate the network topology and external field in \prettyref{fig:RandomCoeffiicients}\textbf{(a)}.

\begin{figure*}
\begin{centering}
\includegraphics{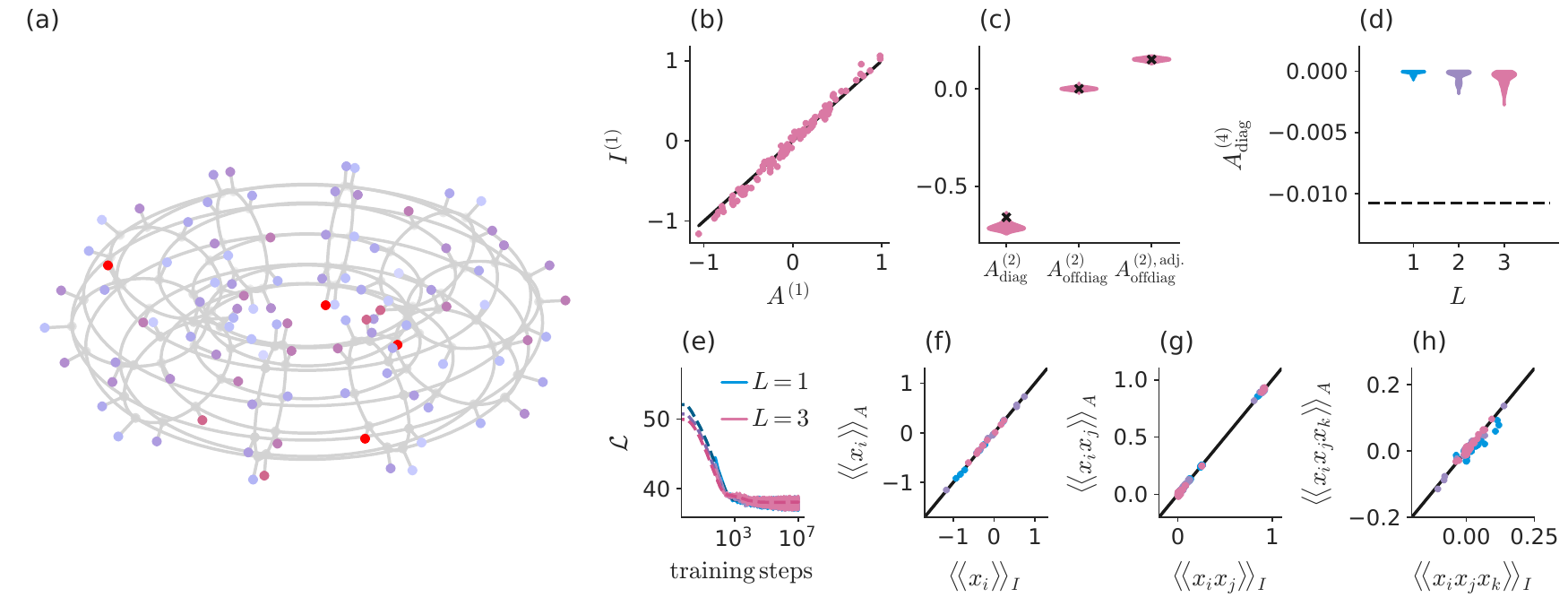}
\par\end{centering}
\caption{\textbf{Symmetry broken lattice model} for networks of varying depth
trained on a $d=10^{2}$\textbf{ }dimensional data set with $D=10^{6}$\textbf{
}samples. \textbf{(a) }Sites of square lattice with periodic boundary
conditions distributed on a two-dimensional torus. Colored dots show
strength of external field $h$ at connected lattice sites. \textbf{(b)}
Learned over true first order coefficients for network depth $L=3$.
\textbf{(c) }Distribution of learned coefficient entries $A^{(2)}$
compared to target values (black crosses) for network depth $L=3$.
We distinguish self-interaction terms $A_{\mathrm{diag}}^{(2)}$ from
off-diagonal entries $A_{\mathrm{offdiag}}^{(2)}$. From the off-diagonal
entries $A_{\mathrm{offdiag}}^{(2)}$, we further separate those entries
belonging to adjacent lattice sites $A_{\mathrm{offdiag}}^{(2),\mathrm{adj.}}$.
\textbf{(d) }Training loss (solid curves) and test loss (dashed curves).
Colors distinguish different network depths $L$. \textbf{(e) }Distribution
of learned fourth order self-interactions as function of network depth.
The dashed line marks the target value. \textbf{(f--h) }Learned over
true cumulants of up to third order. Cumulants were computed on a
subset of $10$ randomly chosen lattice sites. Colors distinguish
different network depths $L$.\label{fig:Coefficients-of-lattice-model}}
\end{figure*}

As in \prettyref{subsec:Out-of-class-distributions}, we sample from
this distribution using an MCMC sampler (see \prettyref{app:Sampling-random-actions}
for details) and train networks of different depths $L$. In \prettyref{fig:Coefficients-of-lattice-model}
\textbf{(b-d)} we compare the learned action coefficients $A^{(k)}$
to the corresponding target values $I^{(k)}.$ We find good agreement
for $A^{(1)}$ and the off-diagonal values $A_{ij}^{(2)},i\neq j$,
independent of network depth. The external field and nearest-neighbor
coupling is therefore recovered accurately. The self-interaction $A_{\mathrm{diag}}^{(2)}$
is typically lower than the target value while the fourth-order self-interaction
$A_{\mathrm{diag}}^{(4)}$ is typically larger. Since both parameters
control the widths of the distributions, the slightly lower $A_{\mathrm{diag}}^{(2)}$
can compensate for the too small magnitude of $A_{\mathrm{diag}}^{(4)}$,
producing an effective theory.
Nevertheless the higher-order coefficients $A^{(k\geq3)}$ of the
learned theory are relevant. We show in \prettyref{fig:Coefficients-of-lattice-model}
\textbf{(f-h)}, that the first, second, and third cumulants of the
learned and true distribution approach each other as the network depth
increases. A Gaussian approximation of $S_{I}$ would only tune $A^{(1)}$
and $A^{(2)}$ with $A^{(k\geq3)}=0$ to match the first and second
cumulants shown in \prettyref{fig:Coefficients-of-lattice-model}
\textbf{(f,g)}. As in the multimodal case therefore, we learn an effective
non-Gaussian theory. The effective theory may result from the depth
of the network being small in comparison to the dimensionality of
the system to tune all higher order coefficients $A^{(k\geq3)}$.
The number of independent entries in the action coefficients up to
the fourth order is $\mathcal{O}\left(d^{4}\right)$, roughly $4.6\cdot10^{6}$
for the case of $d=10^{2}$, with by far the most number of entries
in the highest-order coefficient $A^{(4)}.$ In contrast, the number
of free parameters of a single layer is $\left\lceil \frac{d}{2}\right\rceil ^{3}$
in $\tilde{\chi_{l}}$, $d^{2}$ in $W_{l}$, and $d$ in \textbf{$b_{l}$},
so all in all $\left\lceil \frac{d}{2}\right\rceil ^{3}+d(d+1)$.
Although the coefficients do not depend linearly on the network parameters,
this gives a rough estimate of the required depth $L=34$ of the network,
at which the number of free parameters in the network and in the coefficients
coincide. Since the number of entries in the coefficients grows with
$\mathcal{O}\left(d^{4}\right)$, but the number of free parameters
in the network is only $\mathcal{O}\left(Ld^{3}\right)$, the depth
required to tune all coefficient entries grows with $d$. Below this
depth, it may well be that the flexibility of the network is too small
to tune all fourth-order action coefficients. Even though $\bigl\langle\left(x^{\otimes k}\right)_{\alpha}\bigr\rangle_{\mathcal{D}}-\bigl\langle\left(x^{\otimes k}\right)_{\alpha}\bigr\rangle_{A}$
in \prettyref{eq:noisy_update_coefficients} is likely non-zero then,
the coefficients may still reach stationary values by the combination of the terms on the right hand side of \prettyref{eq:noisy_update_coefficients}
vanishing. Indeed we find in \prettyref{app:Lattice-model-in-low-dimensions},
that in lower dimensions smaller network depths are sufficient to
tune the higher statistical orders. Furthermore, the learned coefficients
$A^{(4)}$ approach the target value as the depth increases. In all
examples studied here, the alignment between learned and true statistics
improves with depth, which increases the network flexibility. For
shallow networks, however, although the learned action is not equal
to the true one, it effectively describes the statistics of the true
distribution beyond the Gaussian order as indicated by the good agreement
of the cumulants. This behavior is equivalent to that of renormalized
theories \citep{Amit84}, which feature the same statistical correlations
while changing the interaction strengths in a consistent manner.

\section{Discussion\label{sec:discussion}}

We have developed a method to learn a microscopic theory from data
-- concretely, we learn a classical action that assigns a probability
to each observed state. For this data-driven approach, we employ a
specific class of deep neural networks that are invertible and that
can  be trained in an unsupervised manner, without the need of labeled
training data. Such networks have been used before as generative models
\citep{DInh15_1410,dinhDensityEstimationUsing2017,kingmaGlowGenerativeFlow2018a}
 but are generally considered a black box: after training the learned
information is stored in a large number of parameters in an accessible,
yet distributed and generally incomprehensible manner.

The diagrammatic formalism developed here allows us to extract the
data statistics from the trained network in terms of an underlying
set of interactions -- a common formulation used throughout physics.
To achieve this, we designed the network architecture as a trade-off
between flexibility and analytical tractability. The choice of a quadratic
polynomial, along with a volume-preserving invertible architecture,
allows us to obtain explicit expressions for the interaction coefficients.
 This formalism shows how the interplay between linear and nonlinear
mappings in the network composes non-Gaussian statistics, and hence
higher-order interactions, in a hierarchical manner. As a consequence
of the quadratic interaction constituting the fundamental building
block, higher-order interactions are decomposed into this simplest
possible form of nonlinear interplay. As a result, the order of interaction
in the data directly maps to the required depth of the network in
an understandable manner, thus providing an explanation why deep networks
are required to learn higher-order interactions.

The analytical framework we developed can also readily be extended
to higher-order nonlinearities: In terms of the diagrammatic language,
the quadratic activations used in this work amount to splitting of
``legs'' in the Feynman diagrams into pairs. Likewise one obtains
a three-fold splitting from a cubic term, a four-fold splitting from
a quartic term and so on. Such higher-order nonlinearities would allow
the building of more complex interactions with fewer layers.

From a physics point of view one may regard the trained network as
a device to solve an interacting classical field theory in a data-driven
manner: once the network has been trained, it maps each configuration
of the interacting theory in data space to samples in latent space
that follow a Gaussian theory, hence a non-interacting one. Such
a mapping allows one to compute arbitrary connected correlation functions
of the interacting theory. The framework offers two routes to this
end. The traditional one uses common rules of diagrammatic perturbation
theory to obtain controlled approximations of connected correlation
functions in terms of connected diagrams constructed from propagators
and interaction vertices of the inferred action. An alternative one
directly constructs connected correlation functions hierarchically
across the layers of the network, ultimately reduced to pairwise interactions
on the level of the latent Gaussian. For example, the second order
correlations read $\langle\!\langle x_{i}x_{j}\rangle\!\rangle_{p_{\theta}}=\langle\!\langle f_{\theta,i}^{-1}(z)f_{\theta,j}^{-1}(z)\rangle\!\rangle_{z\sim\mathcal{N}(0,\mathds{1})}$.
For the n-th order correlations $\langle\!\langle x_{i_{1}}\cdots x_{n}\rangle\!\rangle_{p_{\theta}}$
one can therefore either work out the coefficients of the polynomial
$f_{\theta,i_{1}}^{-1}(z)\cdots f_{\theta,i_{n}}^{-1}(z)$, in a similar
manner to the action transform, and then average the resulting function
over $p_{Z},$ or estimate the correlations by drawing samples from
the generative network. An open avenue to explore further in this
regard is the link between the presented framework and asymptotically
free theories, where an interacting theory becomes non-interacting
at high energy (UV) scales. In that case, different scales are connected
by a renormalization group (RG) flow.  It would be interesting to
investigate whether the change of couplings described by the RG flow
can be related to the transformations performed by the network. More
broadly, the ability to learn an interacting theory by the network
can be considered an alternative to asymptotic freedom, as the flow
across layers does not have to correspond to a change of length scale.

To provide the most transparent setting, we have here chosen the simplest
but common case of a latent Gaussian distribution, which has the
aforementioned advantage of mapping to a non-interacting theory. A
consequence is that the latent distribution only has a single maximum.
Since for invertible volume-preserving networks the number of modes
in data space and in latent space are identical, these network architectures
therefore only learn monomodal distributions. For many settings of
interest this is sufficient: multi-modal distributions in physical
systems often occur together with non-ergodic behavior such as spontaneous
symmetry breaking, selecting one of the modes of the distribution.
As presented, our approach necessarily learns the statistics of the
selected mode, and thus obtains an effective theory of the single
selected phase of the system. The simplest way to learn genuine multi-modal
distributions is the use of a multi-modal latent distribution, such
as a Gaussian mixture. Our framework would then provide one set of
action coefficient for each mixture component, correspondingly offering
one effective theory for each phase.

We complement this study with a characterization of the training process,
confirming the expectation that both larger training set size and
network depth improve learning. Larger data sets in general decrease
the bias of the learned distribution due to undersampling of the true
distribution. This point is most severe for higher-order statistics,
while the first two orders of the statistics are typically learned
robustly also from limited data. We provide an approximate expression,
\prettyref{eq:noisy_update_coefficients}, to investigate the convergence
properties of statistics of different orders. While deeper networks
are required to offer sufficient flexibility to learn higher-order
statistics, the larger number of trainable parameters at the same
time requires more to learn the statistics accurately. Alternatively,
the network flexibility can be increased by raising the order of the
polynomial activation function. Finding the optimal tradeoff between
local nonlinearity and depth is an interesting point of future research.

Several studies have highlighted the role of the stochasticity of
the training algorithm for networks performing classification \citep{liStochasticModifiedEquations2017,chaudhariStochasticGradientDescent2018a,mignaccoDynamicalMeanfieldTheory2021}.
A possible starting point to understanding SGD for generative models
could be \prettyref{eq:noisy_update_coefficients}, which relates
the trajectory of the learned distribution in coefficient space to
the training algorithm. Equation \eqref{eq:noisy_update_coefficients}
closely resembles the training of Restricted Boltzmann machines (RBMs),
where the pairwise coupling matrix between hidden and visible layers
updated according to the difference between learned and observed pairwise
correlations \citep{ackleyLearningAlgorithmBoltzmann1985}. Studying
\prettyref{eq:noisy_update_coefficients} could shed light on the
dynamics of unsupervised learning, for which the architecture used
in this work is a fully tractable prototype.

Another challenge in learning interacting theories of higher order
is the necessarily large size of the action coefficients which grows
with the dimension, irrespective of the manner in which these interactions
are inferred. However, it is plausible that not all terms in these
tensors are equally important: For spatially or temporally extended
systems, interactions between distant degrees of freedom may be irrelevant.
The framework explicitly shows how higher-order interactions are composed
out of lower-rank coefficients. This may be leveraged to extract the
most relevant contributions in a tractable manner (see \prettyref{app:decomposed_tensors}).

The general problem of inferring models from data discussed in this
work is a well-known challenge. In the dynamical systems setting,
the authors of \citep{bruntonDiscoveringGoverningEquations2016,casadiegoModelfreeInferenceDirect2017,millerLearningTheoryInferring2020,lu_learning_2019}
use regression to infer the right hand side of the governing differential
equation of a system from a set of basis functions. Other studies
\citep{jordanEvolvingInterpretablePlasticity2021,confavreuxMetalearningApproachRe2020}
infer rules for the time-dependence of couplings (synaptic plasticity)
using regression and genetic programming. These approaches produce
interpretable models, but require a pre-determined a set of basis
functions or operations, through the combination of which the system
dynamics are approximated. Inference of parameters stochastic processes
\citep{opper_variational_2007,opper_variational_2019,vrettas_variational_2015,ruttor_efficient_2009}
also relies on the specific form of the update equations. Prior knowledge
about likely terms in the dynamical equations or their exact functional
form is therefore needed in these works.

Many approaches exist to infer pair-wise couplings of binary degrees
of freedom, such as for Ising models \citep{Amit84}. Prominent among
them are Boltzmann machines \citep{ackleyLearningAlgorithmBoltzmann1985}.
Other techniques for Ising models first solve the forward problem,
namely the statistics given the couplings -- using variations of
mean-field theory \citep{mezardExactMeanField2011,schneidmanWeakPairwiseCorrelations2006,zengNetworkInferenceUsing2011}
or the TAP equations \citep{Thouless77_593,Vasiliev74,sakellariouEffectCouplingAsymmetry2012}
-- and then invert these relations explicitly or iteratively \citep{Zeng11_041135,Roudi11,baldassiInverseProblemsLearning2018,sakellariouEffectCouplingAsymmetry2012}.
Maximum likelihood methods or the TAP equations have also been used
to infer the patterns of Hopfield models \citep{decelleInverseProblemsStructured2021,coccoHighdimensionalInferenceGeneralized2011}.
In the special case of a tree-like, known network topology, or translationally
invariant higher-order couplings along a linear chain, the inverse
problem can be solved exactly \citep{mastromatteoInverseIsingModel2012}.
Further works maximize the likelihood of the network model given the
data, by using belief propagation to reconstruct the network structure
from infection cascades \citep{braunsteinNetworkReconstructionInfection2019},
or Monte Carlo sampling to infer amino acid sequences in proteins
\citep{moraMaximumEntropyModels2010}. In contrast to these works,
in this study we consider continuous rather than discrete variables.
Furthermore, we are not restricted to pairwise interactions and do
not require prior knowledge on the structure of interactions.

\citet{zacheExtractingFieldTheory2020} also solve the forward problem:
they approximate a higher-order interacting theory to tree-level or
one-loop-order in the effective action, and thereby obtain an invertible
relation between interactions and correlations. This approach relies
on the validity of the approximations, namely for the typically difficult
step from interactions to correlation functions. These approximations
are not necessary to train INNs, as the correlation functions are
implicitly generated by the network mapping.

Neural networks have also previously been used to treat inverse problems.
They are trained to infer the posterior probability of characteristic
parameters given data \citep{goncalvesTrainingDeepNeural2020,ardizzoneAnalyzingInverseProblems2019},
and to compute renormalized degrees of freedom that are maximally
informative about the global state of a system \citep{gokmenStatisticalPhysicsLens2021}.
For systems of interacting identical particles, \citet{NEURIPS2020_c9f2f917}
use symbolic regression on trained Graphical Neural Networks to derive
interpretable interactions. However, these approaches make no lucid
connection between the learned model and the parameters of the neural
network as we do in this study.

With the here proposed extraction method for the action of a physical
system at hand, one can now proceed to extract hitherto unknown interacting
models from data. One interesting application of this approach is
to learn actions for systems for which a microscopic or mesoscopic
description is not known, for example in biological neuronal networks:
the inferred coefficients would determine the importance of nonlinear
interactions in biological information processing. The approach may
also be fruitful when applied to systems in physics where the microscopic
theory may be known but an effective theory is sought that captures
an observed macroscopic phenomenon.

\begin{acknowledgments}
We are grateful to Thorben Finke, Sebastian Goldt, Christian Keup,
Michael Krämer and Alexander Mück for helpful discussions. This work
was partly supported by the German Federal Ministry for Education
and Research (BMBF Grant 01IS19077A to Jülich and BMBF Grant 01IS19077B
to Aachen) and funded by the Deutsche Forschungsgemeinschaft (DFG,
German Research Foundation) - 368482240/GRK2416, the Excellence Initiative
of the German federal and state governments (ERS PF-JARA-SDS005),
and the Helmholtz Association Initiative and Networking Fund under
project number SO-092 (Advanced Computing Architectures, ACA).\textit{
}Open access publication funded by the Deutsche Forschungsgemeinschaft
(DFG, German Research Foundation) -- 491111487.
\end{acknowledgments}

\appendix

\section{Diagrammatic update equations\label{app:diagrams}}

We here describe the diagrammatic rules to compute the action coefficient
transforms in Eqs.~\eqref{eq:coefficient_update_linear} and \eqref{eq:coefficient_update_nonlinear}.

{\small{}}Following the structure in \prettyref{sec:Learning-Actions-with-INNs},
we first treat the linear transform \eqref{eq:linear_transform}.
Each action coefficient $A_{l+1}^{(k)}$ is represented by a vertex,
where the number of outgoing lines, also called legs, is equal to
the rank of the coefficient. Each leg is assigned an index $i_{l}$
corresponding to the indices $(i_{1},\dots,i_{k})$ of the coefficient
entry of $A_{l+1}^{(k)}$ it represents. Equation \eqref{eq:coefficient_update_linear}
shows that each index of the coefficient must either be contracted
with $W_{l}$, and therefore remains a leg of the resulting vertex
from the second index of $W_{l}$, or it must be contracted with $b_{l},$and
therefore drops out. We represent the contraction with $W_{l}$ by
an elongated line decorated with $W_{l}$, and the contraction with
$b_{l}$ as a leg ending on an empty circle. Thus to compute the transformed
action coefficients, we must add empty circles to the previous vertices
in all possible ways and then elongate the remaining legs using $W_{l}$.
A diagram with $k$ legs therefore produces the following diagrams:

\begin{align*}
\vcenter{\hbox{\includegraphics[scale=1]{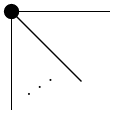}}}\rightarrow &  \vcenter{\hbox{\includegraphics[scale=1]{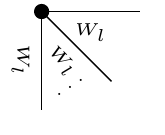}}} 
,\binom{k}{1} \vcenter{\hbox{\includegraphics[scale=1]{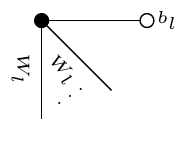}}}, \\
& \dots ,\binom{k}{k} \vcenter{\hbox{\includegraphics[scale=1]{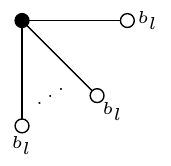}}}
\end{align*}where the combinatorial factors $\binom{k}{l}$ arise due to the
different ways of choosing legs which are contracted with $b_{l}$.
For example, there are $\binom{k}{1}=k$ ways of choosing a single
leg from a vertex with $k$ legs which is then contracted with \textbf{$b_{l}$}.
We first compute the new diagrams for all $k$, then sum up all diagrams
that have equal numbers of legs to one coefficient. This illustrates
how higher-order action coefficients, by the contraction of their
indices with the biases, i.e. the attachment of empty circles to their
legs, contribute to lower-order coefficients.

For the nonlinear transform, we have the opposite effect: each index
in the coefficient either remains or is contracted with $\chi_{l}$,
which increases the rank of the transformed diagram by one. Therefore
either the legs of vertices remain as they are, or they must be split
into two to signify the contraction with $\chi_{l}$, which increases
the number of legs of the vertex by one. We keep the split legs distinguishable
from threepoint vertices by using curved lines for the split legs
and sum over all possible ways to split legs. A diagram with $k$
legs therefore produces the following diagrams:

\begin{align*}
\vcenter{\hbox{\includegraphics[scale=1]{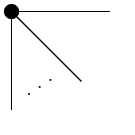}}} \rightarrow &
\vcenter{\hbox{\includegraphics[scale=1]{Figures/Diagrams/NonLinearTransformLHS.pdf}}} , 
\binom{k}{1} \vcenter{\hbox{\includegraphics[scale=1]{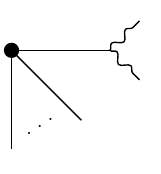}}} \\
& \dots , \binom{k}{k} \vcenter{\hbox{\includegraphics[scale=1]{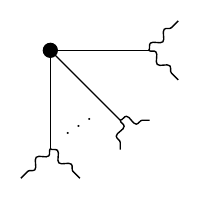}}} 
\end{align*}

Here, the combinatorial factors arise due to the number of ways in
which to choose the split legs. The number of factors $\chi_{l}$
in any diagram can then be read off from the number of leg splits.
As in the linear transform, to compute the transformed action coefficients
of rank $k,$ we must therefore sum over all diagrams with equal numbers
of legs. This illustrates how higher-order action coefficients arise
through the splitting of legs by factors of $\chi_{l}$.

\section{Decomposed tensors\label{app:decomposed_tensors}}

\begin{figure}
\begin{centering}
\includegraphics{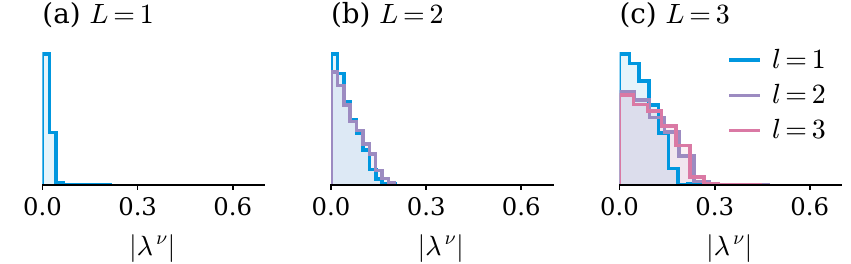}
\par\end{centering}
\caption{\textbf{Eigenvalue distributions of decomposed $\chi_{l}$ }for networks
of different depths. We decompose trained network parameters $\chi_{l}$
from \prettyref{subsec:Interaction-on-a-lattice} to the form of
\prettyref{eq:chi_decomposed} and distinguish eigenvalues from the
decomposed form of different layers $l$. \label{fig:chi_eigenvalues}}
\end{figure}

Higher-order tensors $T^{(k)}$ of rank $k$ can become numerically
intractable for large dimension $d$ as the number of entries in $T^{(k)}$
grows as $\mathcal{O}\left(d^{k}\right)$. Specifically, two challenges
arise: First, to store the entries of the tensors. Second, to compute
contractions with matrices $W$ such as 
\begin{equation}
T^{(k)}\cdot\left(W\right)^{\otimes k}\,,\label{eq:tensor_matrix_contraction}
\end{equation}
which arise due to the linear coefficient transform \prettyref{eq:coefficient_update_linear}.
For these contractions, we must compute the sum over all entries
\[
\left(T^{(k)}\cdot\left(W\right)^{\otimes k}\right)_{i_{^{1}}^{.},\dots i_{k}}=\sum_{j_{1},\dots j_{k}=1}^{d}T_{j_{1},\dots j_{k}}^{(k)}W_{j_{1},i_{1}}\cdots W_{j_{k},i_{k}}\,;
\]
therefore without further simplification this entails the computation
of $d^{k}$ entries of $T^{(k)}\cdot\left(W\right)^{\otimes k}$ from
$d^{k}$ terms each, so the total number of floating point operations
scales as $\mathcal{O}(d^{2k})$. Even though the number of steps
required therefore only grows polynomially with $d$, for realistic
data set sizes and $k=4$, this number increases very fast.

To facilitate the computation of coefficients with rank $k=4$, we
exploit that they are built from coefficients of lower rank to write
the tensors in a decomposed form. 

As a first step, we decompose the network parameters $\chi_{l}$.
Without loss of network expressivity, we may choose $\chi_{l}$ to
be symmetric in its latter two indices $\left(\chi_{l}\right)_{\mu jk}=\left(\chi_{l}\right)_{\mu kj}$.
In the following, we drop the layer index $l$ for brevity, as the
structure of the computation is the same for any layer. We then rearrange
the tensor to be a list of $d$ symmetric matrices $\bar{\beta}^{\mu},\,\mu=1,\dots,d$
such that $\chi_{\mu kj}=\bar{\beta}_{kj}^{\mu}$. Using the eigendecomposition
of these matrices $\bar{\beta}^{\mu}$, we may write

\begin{align}
\chi & =\sum_{\mu,\nu=1}^{d}\gamma^{\mu,\nu}\otimes\beta^{\mu,\nu}\otimes\beta^{\mu,\nu}\label{eq:chi_decomposed}
\end{align}
where $\gamma^{\mu,\nu},\beta^{\nu}$ are vectors and $\gamma_{\tau}^{\mu,\nu}=\delta_{\tau,\mu}\lambda_{\nu}^{\mu}$
has only one non-zero entry, namely the $\nu$-th eigenvalue of the
$\mu$-th matrix $\bar{\beta}^{\mu}$. To store this object we require
$2d^{2}$ vectors of length $d$, namely $d^{2}$ vectors $\beta^{\mu,\nu}$
and $d^{2}$ vectors $\gamma^{\mu,\nu}$.The magnitude of entries
in $\chi$ is directly related to the magnitude of the eigenvalues
$\lambda_{\nu}^{\mu}$, which is typically small for trained networks.
We show distributions of eigenvalues from trained networks in \prettyref{fig:chi_eigenvalues}.
The distributions broaden with increasing depth, however the peak
of the distribution remains at $\left|\lambda_{\nu}^{\mu}\right|=0$.
It is therefore possible to reduce the space required to store $\chi$
and all tensors related to it by placing a cutoff $\bar{\lambda}\geq\text{0}$
on the eigenvalues, keeping only the $\bar{n}\leq d^{2}$ largest
eigenvalues which have $\left|\lambda_{\nu}^{\mu}\right|\geq\text{\ensuremath{\bar{\lambda}}}$.
Then the number of entries required to store $\chi$ scales as $\mathcal{O}\left(2\bar{n}d\right)$,
as again we need $2\bar{n}$ vectors of length $d$ each. To further
simplify the expression, we absorb the sum over $\mu,\,\nu$ into
a single index $\tau=1,\dots,\bar{n}$.

An alternative way to achieve the decomposition of $\chi$ into a
reduced number of components would be to use the decomposed form \prettyref{eq:chi_decomposed}
directly during training and limit the number of independent vectors
$\beta^{\tau}$. This approach effectively trades the network expressivity
for the tractability of the action coefficient transforms.

In addition to reduced storage requirements,  the decomposed form
\eqref{eq:chi_decomposed} also has the advantage is that the decomposition
translates to all tensors computed via contraction with $\chi$, which
is how higher order coefficients are originally generated (compare
with \prettyref{eq:coefficient_update_nonlinear}). The contraction
between a rank $k$ symmetric tensor $T^{(k)}$ and $\chi$ is

\[
T^{(k)}\cdot\chi=\sum_{\tau}\left(T^{(k)}\cdot\gamma^{\tau}\right)\otimes\beta^{\tau}\otimes\beta^{\tau}\,.
\]

If $k=1$, the result is just a matrix. If $k=2$, then $T^{(k)}\cdot\gamma^{\tau}=:\alpha^{\tau}$
is a vector, therefore $T^{(k)}\cdot\chi$ can be written as a sum
of outer products between three vectors. If $k=3,$ the result is
a sum of outer products between a matrix $T^{(k)}\cdot\gamma^{\tau}=\bar{\alpha}^{\tau}$
and two vectors;
\begin{equation}
T^{(3)}\cdot\chi=\sum_{\tau}\bar{\alpha}^{\tau}\otimes\beta^{\tau}\otimes\beta^{\tau}\,.\label{eq:decomposed_contraction_T3_chi}
\end{equation}
 The matrices$\bar{\alpha}^{\tau}$ are symmetric since
\[
\bar{\alpha}_{ab}^{\tau}=\sum_{c}T_{abc}^{(k)}\gamma_{c}^{\tau}=\sum_{c}T_{bac}^{(k)}\gamma_{c}^{\tau}=\bar{\alpha}_{ba}^{\tau}\,.
\]
The case $k\geq4$ does not arise, as any coefficient with degree
$k\geq4$ must already contain at least two factors $\chi$.

To store the factors of \prettyref{eq:decomposed_contraction_T3_chi}we
therefore require $\bar{n}$ matrices $\text{\ensuremath{\bar{\alpha}^{\tau}}}$
and $\bar{n}$ vectors $\beta^{\tau}$. The number of matrix and vector
entries required to store this object is therefore $\mathcal{O}\left(\bar{n}d^{2}\right)$.
The contraction with matrices $W$ along all indices is 
\[
\left(T^{(3)}\cdot\chi\right)\cdot\left(W\right)^{\otimes4}=\sum_{\tau}\left(W^{T}\bar{\alpha}^{\tau}W\right)\otimes\left(W^{T}\beta^{\tau}\right)\otimes\left(W^{T}\beta^{\tau}\right)\,,
\]
which corresponds to $\bar{n}$ matrix-vector products $W^{T}\beta^{\tau}$
and $2\bar{n}$ matrix-matrix products for $\bar{\alpha}^{\tau}W$
and $W^{T}\left(\bar{\alpha}^{\tau}W\right)$. Each term in the matrix-matrix
product is computed from $d$ terms, therefore the number of terms
required to compute $W^{T}\bar{\alpha}^{\tau}W$ is $\mathcal{O}(2d^{\omega})$
with the matrix multiplication exponent $\omega$, which depends on
the concrete algorithm used for matrix multiplication, e.g. $\omega\approx2.8$
for the Strassen algorithm \citep{strassenGaussianEliminationNot1969}.
To evaluate the contraction, we therefore need to compute $\mathcal{O}($$2\bar{n}d^{\omega})$
terms. Even in the case of no cutoff $\bar{\lambda}=0\Rightarrow\bar{n}=d^{2}$,
this approach significantly reduces the required computations compared
to the naive implementation.

In \citep{schatzExploitingSymmetryTensors2014a} it was shown that
the number of floating point operations needed to compute general
contractions of the type of \prettyref{eq:tensor_matrix_contraction}
can be reduced by exploiting the symmetry of the tensors. They propose
a simple scheme to reduce the number of floating point operations
(using $\omega=3)$ to roughly $\mathcal{O}(d^{k+1}),$ and a more
complex structure of saving these tensors, which further speeds up
the computations at the expense of storing more intermediate entries.
In our experiments, we have used a maximal dimensionality of $d=10^{2}$
in \prettyref{subsec:Interaction-on-a-lattice} and no cutoff $\overline{\lambda}=0.$
In the absence of any cutoff, we find a scaling of our algorithm roughly
equal to the simpler scheme proposed in \citep{schatzExploitingSymmetryTensors2014a}.
The combination of a cutoff, more efficient storing of symmetric tensors
as suggested in \citep{schatzExploitingSymmetryTensors2014a}, or
restricting the number of free components in $\chi$ directly, facilitates
the extension of the coefficient transforms to higher dimension $d$.

\section{Dissimilarity of parameters\label{app:Dissimilarity-of-parameters}}

\begin{figure}
\begin{centering}
\includegraphics{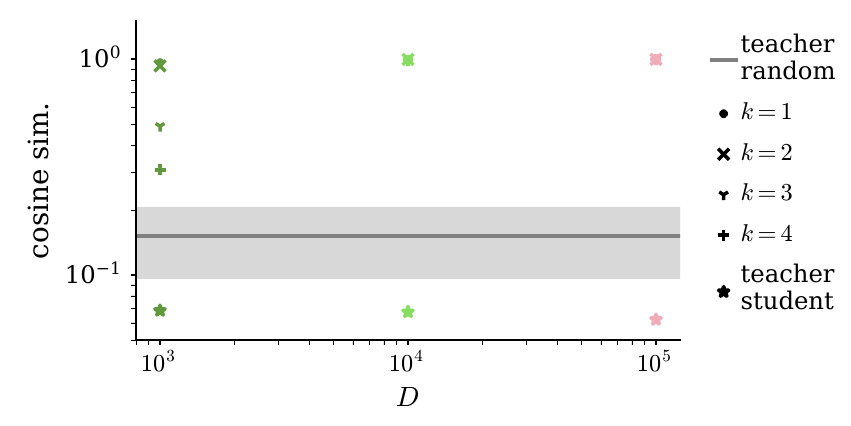}
\par\end{centering}
\caption{\textbf{Dissimilarity of parameters.} Stars show the cosine similarities
of the teacher and student network parameters trained on varying data
set sizes $D$. The average cosine similarity between the teacher
and $10^{2}$ randomly generated random networks is marked by the
grey line, the shaded area encompasses one standard deviation. The
remaining markers display the cosine similarity between the teacher
coefficients $T^{(k)}$ and student coefficients $S^{(k)}.$\label{fig:Dissimilarity-of-parameters.}}
\end{figure}

We here show that although the learned statistics of two networks
may be the same, their parameters do not need to align. To this end,
we use the teacher-student setup presented in \prettyref{subsec:In-class-distributions},
and compute the cosine similarities between pairs of network parameters
$b_{l},\,W_{l},\,\chi_{l}$ -- both between the teacher and the student,
and between the teacher and a network of the same architecture with
random Gaussian weights. We then average over the cosine similarities
of the different parameters to obtain the average network cosine similarity.
We show in \prettyref{fig:Dissimilarity-of-parameters.} that although
the teacher and student coefficients approach each other for increasing
data set sizes $D$, their network parameters remain dissimilar. The
appropriate object to compare the learned statistics is therefore
the action, not the parameters of the network.

\section{Learning action coefficients with SGD \label{app:Coefficient-update-equation}}

The learned action $S_{\theta}(x)$ depends on the parameters $\theta$
only through the coefficients $A^{(k)}$. We can therefore also
view the training as a nonlinear optimization of the action coefficients
via the parameters. However, we cannot freely move in this coefficient
space: we must ensure that the action stays normalized, $\int\exp(S_{\theta}(x))dx=1$.
To this end we fix the constant term in the action: 
\begin{equation}
A^{(0)}=-\ln\int\exp\left(\sum_{k\geq1}A^{(k)}\cdot\left(x\right)^{\otimes k}\right)dx\,.\label{eq:normalization_condition}
\end{equation}

Given this constraint, we rewrite the update equation \eqref{eq:parameter_updates}
in terms of the coefficients using the chain rule,

\[
\Delta\theta_{i}=\eta\sum_{k\geq1}\sum_{\alpha_{k}}\frac{\partial}{\partial A_{\alpha_{k}}^{(k)}}\langle S_{\theta}(x)\rangle_{\mathcal{D}_{t}}\frac{\partial A_{\alpha_{k}}^{(k)}}{\partial\theta_{i}}\,,
\]
where $\sum_{\alpha_{k}}$ runs over all possible (multi-)indices
of $A^{(k)}.$ The update in the parameters induces a change in the
action coefficients. We use this to approximate the update step for
the coefficient entry $A_{\alpha}^{(k)}$ to linear order in the parameter
updates:

\begin{align}
\Delta A_{\alpha}^{(k)}  \approx &\sum_{i}\frac{\partial A_{\alpha}^{(k)}}{\partial\theta_{i}}\Delta\theta_{i}\nonumber \\
 = &\eta\frac{\partial}{\partial A_{\alpha}^{(k)}}\langle S_{\theta}(x)\rangle_{\mathcal{D}_{t}}\sum_{i}\left(\frac{\partial A_{\alpha}^{(k)}}{\partial\theta_{i}}\right)^{2} \nonumber\\
 & + \eta \sum_{l,\alpha_l \neq \alpha} \frac{\partial}{\partial A_{\alpha_l}^{(l)}}\langle S_{\theta}(x)\rangle_{\mathcal{D}_{t}}\sum_{i} \frac{\partial A_{\alpha}^{(k)}}{\partial\theta_{i}}\frac{\partial A_{\alpha_l}^{(l)}}{\partial\theta_{i}} 
 \,.\label{eq:coefficient_update_SGD}
\end{align}
The update step $\Delta A_{\alpha}^{(k)}$ therefore depends on the
network architecture through the derivatives $\frac{\partial A_{\alpha}^{(k)}}{\partial\theta_{i}}$.
The derivative of the expectation of the action with respect to the
the coefficients in the former factor in \prettyref{eq:coefficient_update_SGD}
is
\begin{align}
\frac{\partial}{\partial A_{\alpha}^{(k)}}\langle S_{\theta}(x)\rangle_{\mathcal{D}_{t}} & =\langle\left(x^{\otimes k}\right)_{\alpha}\rangle_{\mathcal{D}_{t}}+\frac{\partial A^{(0)}}{\partial A_{\alpha}^{(k)}}\nonumber \\
 & =\langle\left(x^{\otimes k}\right)_{\alpha}\rangle_{\mathcal{D}_{t}}-\langle\left(x^{\otimes k}\right)_{\underline{\alpha}}\rangle_{A},\label{eq:Action_deriv_coefficient}
\end{align}
where $\langle\cdot\rangle_{A}$ denotes the current average of the
learned distribution and we used \prettyref{eq:normalization_condition}
in the second line. This term induces a variability in the coefficient
updates, as $\langle x^{\otimes k}\rangle_{\mathcal{D}_{t}}$ will
vary from batch to batch due to the finite size of each batch. We
define the random variable $\xi_{t}^{(k)}=\langle x^{\otimes k}\rangle_{\mathcal{D}_{t}}-\langle x^{\otimes k}\rangle_{\mathcal{D}}$
to be the deviation between the moment estimated on the current batch
$\mathcal{D}_{t}$ and the moment estimated on the entire data set
$\mathcal{D}$. Combining Eqs.~\eqref{eq:Action_deriv_coefficient}
and \eqref{eq:coefficient_update_SGD} yields \prettyref{eq:noisy_update_coefficients}.

\section{Random generation of multimodal actions\eqref{app:generating-random-actions}\label{app:generating-random-actions}}

\subsubsection*{Coefficient distributions for random actions}

In section \prettyref{subsec:Out-of-class-distributions} we use multimodal
actions $S_{R}$ constructed from randomly drawn coefficients $R^{(k)}$.
A basic condition these coefficients must satisfy is that the resulting
action be normalizable: $\int S_{R}(x)\,dx<\infty$. We note that
for large enough $x$, the action is dominated by the highest order
terms: 
\[
S(x)\xrightarrow[\lVert x\rVert\to\infty]{}\bigl(R^{(4)}\bigr)\cdot x^{\otimes4}\,.
\]
It is therefore necessary and sufficient for normalizability that
$R^{(4)}$ be negative definite, which we ensure by choosing $R^{(4)}$
to be a diagonal tensor with negative coefficients:
\begin{equation}
R_{i_{1}i_{2}i_{3}i_{4}}^{(4)}=\begin{cases}
-\frac{x_{r}^{-4}}{d} & \text{if }i_{1}=i_{2}=i_{3}=i_{4}\,;\\
0 & \text{otherwise}\,.
\end{cases}\label{eq:R4}
\end{equation}
Here $d$ is the dimensionality of the data, and $x_{r}\in\R$ is
a length scale which we are free to choose; one can view $R^{(4)}$
as a regulator term, and $x_{r}$ as the value for which it becomes
strongly suppressing. For our experiments we used $x_{r}=1.0$.

Having ensured that the action is normalizable, we can define the
probability $p_{R}(x\,|\,\lbrace R^{(k)}\rbrace_{k\leq4})=\exp\bigl(S_{R}(x)\bigr)/\int\exp\bigl(S_{R}(x)\bigr)\,dx$
. We choose the $S_{R}$ such that the data can be described as a
perturbation of a Gaussian theory --- we therefore also choose $R^{(2)}$
to be negative definite (i.e. a valid precision matrix). Since any
$d$-dimensional multivariate Gaussian can be written as a linear
combination of $d$ independent Gaussian variables, we define $R^{(2)}$
as follows:

\begin{align}
W_{ij} & \sim\mathcal{N}(0,1/d^{2})\,, & i,j & =1,\dotsc d\,;\nonumber \\
R_{ij}^{(2)} & \coloneqq c-\frac{1}{2}\sum_{a}W_{ia}W_{ja}\,;\label{eq:Second_random_coefficient}
\end{align}
with $c=-0.1$ in our experiments. This is equivalent to transforming
a Gaussian variable $z\sim\mathcal{N}(0,\mathds{1})$ by a linear
transform $x=W^{-1}z$ and then computing the action of $x$ (compare
to \prettyref{eq:linear_mapping_example_A2}).

Finally, the coefficients $R^{(1)}$ and $R^{(3)}$ are chosen as
follows ($i,j,k=1,\dotsc d$):
\begin{align*}
R_{i}^{(1)} & \sim\mathcal{N}\bigl(0,\sigma_{i}^{2}\bigr) & \sigma_{i} & =\frac{x_{r}^{-1}}{d}\,;\\
R_{ijk}^{(3)} & \sim\mathcal{N}\bigl(0,\sigma_{ijk}^{2}\bigr) & \sigma_{ijk} & =\frac{x_{r}^{-3}}{s_{ijk}\gamma_{ijk}}\,.
\end{align*}
The scaling with respect to $x_{r}^{-1}$ and $x_{r}^{-3}$ ensures
that neither linear and cubic terms are negligible within the region
where the regulator term $R^{(4)}$ is non-suppressing. The variable
$\gamma_{\alpha}=\lvert\mathcal{P}(\alpha)\rvert$ in the denominator
of $\sigma_{\alpha}$ is the multiplicity of the index $\alpha$.
This is the number of times the component $R_{ijk}^{(3)}$ appears
in $R^{(3)}$; since coefficients are symmetric, it is equal to the
number of distinct permutations of $(i,j,k)$. We scale the multiplicity
by $s_{\alpha}$, the number of different components which have the
same number of permutations -- for example the permutations of the
indices $(2,1,1)$ and $(5,3,3)$ appear $\text{\ensuremath{\gamma_{ijj}=3}}$
times each, and there are $s_{ijj}=d(d-1)$ distinct entries of such
indices. Scaling $\sigma_{\alpha}$ by $\gamma_{\alpha}$ and $s_{\alpha}$
ensures that both the on-diagonal and off-diagonal components of $R^{(2)},\,R^{(3)}$
are significant.

\subsubsection*{Multimodality of random actions}

\begin{figure}
\begin{centering}
\includegraphics{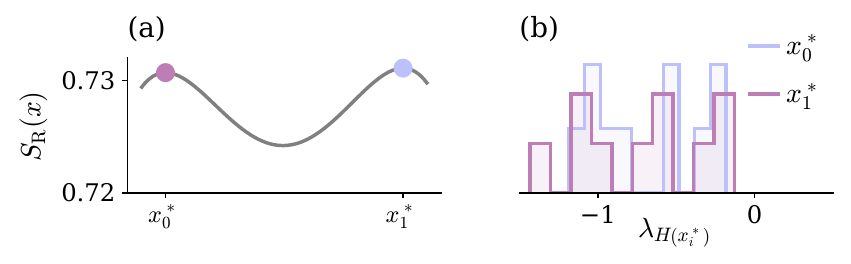}
\par\end{centering}
\caption{\textbf{Multiple local maxima} in $S_{R}$. \textbf{(a) }$S_{R}$
along the straight line connecting two local maxima $x_{0}^{*},\,x_{1}^{*}$
of $S_{R}$ found by the optimization algorithm. \textbf{(b,c) ~}Eigenvalues
of the Hessian~$H$ of $S_{R}$ at local maxima $x_{0}^{*},\,x_{1}^{*}$.
All eigenvalues $\lambda_{H(x_{i}^{*})}$ are negative, therefore
the action is convex down in all directions. \label{fig:Evidence-of-multiple-local-maxima}}
\end{figure}

We here provide evidence that the distribution in \prettyref{subsec:Out-of-class-distributions}
is indeed multimodal. To do so, we initialize an optimization algorithm
at random points and attempt to find the maximum of $S_{R}(x^{*})$.
We initialized at $10^{3}$ different values. In \prettyref{fig:Evidence-of-multiple-local-maxima}
we show the maximal values of $S_{R}(x^{*})$ found by the algorithm
as well as the eigenvalues of the Hessian for selected, distinct final
values $x^{*}$. Since all eigenvalues are below zero, the action
$S_{R}$ is locally convex down in all directions at both points.
The action $S_{R}$ therefore has at least two local maxima.

\section{Sampling actions with MCMC \label{app:Sampling-random-actions}}

Given ground truth coefficients $T^{(k)}$, we create a data set $\mathcal{D}$
by drawing samples according to the probability $p_{C}(x\,|\,\lbrace T^{(k)}\rbrace_{k\leq4})$.
Markov chain Monte Carlo (MCMC) is well suited to this task since
it requires only the unnormalized log-probability, i.e. $S_{C}(x)$,
and is guaranteed to converge to the true distribution (in contrast
to variational methods like ADVI \citep{kucukelbirAutomaticDifferentiationVariational2017}).
To generate $\mathcal{D}$, we used the No-U-Turn Sampler (NUTS) \citep{hoffmanNoUturnSamplerAdaptively2014}
implementation provided by PyMC3 \citep{salvatierProbabilisticProgrammingPython2016};
sampler parameters mostly followed recommended defaults, with $10^{3}$
tuning steps and a mass matrix initialized to unity. The target acceptance
rate was increased to $0.95$ to increase the sensitivity to small
features of the probability distribution. 

\section{Lattice model in low dimensions\label{app:Lattice-model-in-low-dimensions}}

\begin{figure}[h]
\begin{centering}
\includegraphics{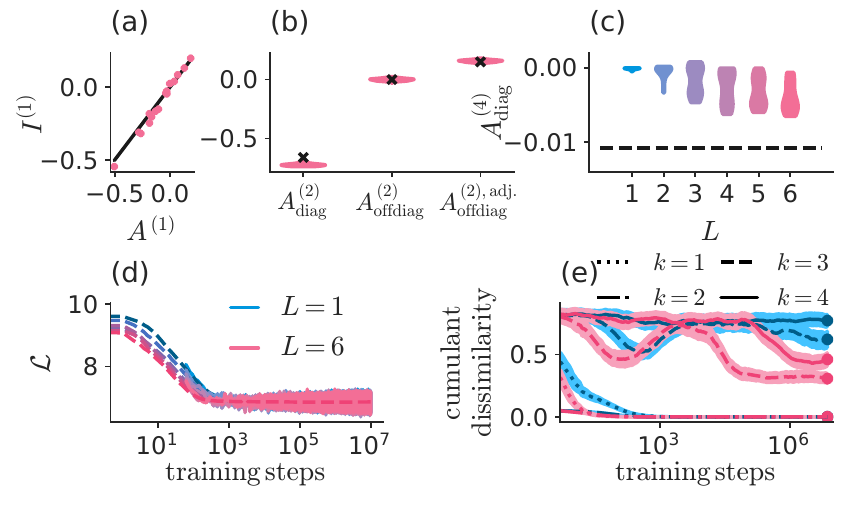}
\par\end{centering}
\caption{\textbf{Coefficients of lattice model} for networks of varying depth
trained on a $d=16$\textbf{ }dimensional data set with $D=10^{5}$\textbf{
}samples. \textbf{(a)} Learned ($L^{(1)})$ over true ($A^{(1)}$)
first order coefficients. \textbf{(b) }Distribution of learned coefficient
entries $A^{(2)}$ compared to target values (black crosses). Self-interaction
terms are labeled $A_{\mathrm{diag}}^{(2)}$, off-diagonal entries
$A_{\mathrm{offdiag}}^{(2)}$. Among the off-diagonal entries $A_{\mathrm{offdiag}}^{(2)}$,
entries belonging to adjacent lattice sites $A_{\mathrm{offdiag}}^{(2),\mathrm{adj.}}$
are shown separately. \textbf{(c) }Training loss (full curves) and
test loss (dashed curves). Colors distinguish different network depths
$L$. \textbf{(d) }Distribution of learned fourth order self-interactions
over network depth. The dashed line marks the target value.\textbf{
(e) }Dissimilarity of true and learned cumulants: $1-\text{\ensuremath{\cos\angle}}\left(\langle\!\langle x^{\otimes k}\rangle\!\rangle_{A},\langle\!\langle x^{\otimes k}\rangle\!\rangle_{R}\right)$
over training steps. We record the cumulants at logarithmically spaced
intervals during training. The curves are smoothed by averaging over
ten adjacent recording steps. Shaded areas show the variation due
to the estimation of the cumulants from samples. Dots indicate training
stage of coefficients shown in (a,b).\label{fig:Coefficients-of-lattice-model-lowdim}}
\end{figure}

We here show a lower dimensional version of the lattice model introduced
in \prettyref{subsec:Interaction-on-a-lattice}.

Here, the combined number of independent entries in the first four
action coefficients is only $4844$, which corresponds to roughly
$6$ network layers. Figure~\ref{fig:Coefficients-of-lattice-model-lowdim}
shows a comparison of true vs. learned coefficients. Independent of
network depth, we find that $A^{(1)}$ and the off-diagonal entries
$A_{ij}^{(2)}$ with $i\neq j$ are recovered correctly (see \prettyref{fig:Coefficients-of-lattice-model-lowdim}~(a,b)).
For shallow networks, the diagonal entries in the fourth order coefficient
$A_{\mathrm{diag}}^{(4)}$ are approximately zero. Their magnitudes
increase with $L$ (see \prettyref{fig:Coefficients-of-lattice-model-lowdim}\textbf{~}(d)).
Increasing the depth $L$ also speeds up learning (see \prettyref{fig:Coefficients-of-lattice-model-lowdim}~(c)).
Furthermore, we find that up to the fourth order, the cumulants of
the learned distribution increasingly align with those of the true
distribution as we increase network depth. Therefore we can conclude
that increasing the depth of the network increases the accuracy of
the learned distribution, both in terms of its coefficients and of
its cumulants. 

\begin{figure}[H]
\begin{centering}
\includegraphics{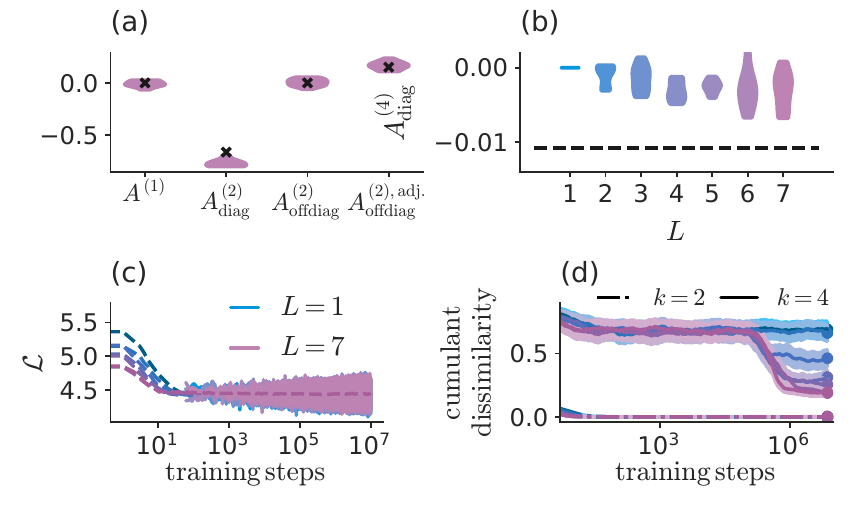}
\par\end{centering}
\caption{\textbf{Coefficients of lattice model without external field} for
networks of varying depth trained on a $d=9$\textbf{ }dimensional
data set with $D=10^{5}$\textbf{ }samples. \textbf{(a) }Distribution
of learned coefficient entries $A^{(1)},\,A^{(2)}$ compared to target
values (black crosses). Self-interaction terms $A_{\mathrm{diag}}^{(2)}$
are shown separately from off-diagonal entries $A_{\mathrm{offdiag}}^{(2)}$.
Among the off-diagonal entries $A_{\mathrm{offdiag}}^{(2)}$, those
entries belonging to adjacent lattice sites $A_{\mathrm{offdiag}}^{(2),\mathrm{adj.}}$
are shown separately. \textbf{(b) }Distribution of learned fourth
order self-interactions compared to network depth. The dashed line
marks the target value.\textbf{ (c) }Training loss (full curves) and
test loss (dashed curves). Colors distinguish different network depths
$L$. \textbf{(d) }Dissimilarity of true and learned cumulants: $1-\text{\ensuremath{\cos\angle}}\left(\langle\!\langle x^{\otimes k}\rangle\!\rangle_{A},\langle\!\langle x^{\otimes k}\rangle\!\rangle_{R}\right)$
over training steps. We record the cumulants at logarithmically spaced
intervals during training. The curves are smoothed by averaging over
ten adjacent recording steps. Shaded areas show the variation due
to the estimation of the cumulants from samples. Dots indicate training
stage of coefficients shown in (a,b).\label{fig:Coefficients-of-lattice-without-h}}
\end{figure}

We repeated the experiment for $d=9$ without any heterogeneous external
field, therefore $h=0$. Again, we find an alignment of most entries
in $A^{(1)},A^{(2)},$with $A_{\mathrm{diag}}^{(2)}$ slightly lower
than expected and $A_{\mathrm{diag}}^{(4)}$ larger than expected
(see \prettyref{fig:Coefficients-of-lattice-without-h}~(a,b)). In
this setting, the symmetry of the action causes the first and third
cumulant to vanish. We therefore only compute the alignment of the
second and fourth cumulants. As in the previous cases, this alignment
increases with depth, as shown in \prettyref{fig:Coefficients-of-lattice-without-h}\textbf{~}(d).
Therefore, the effective nature of the learned theory does not depend
on the system's symmetry being broken.

%\bibliographystyle{apsrev4-2}
%\bibliography{brain,references}
%apsrev4-2.bst 2019-01-14 (MD) hand-edited version of apsrev4-1.bst
%Control: key (0)
%Control: author (8) initials jnrlst
%Control: editor formatted (1) identically to author
%Control: production of article title (0) allowed
%Control: page (0) single
%Control: year (1) truncated
%Control: production of eprint (0) enabled
%

\end{document}